\documentclass[
    aps,
    prb,
    twocolumn,
    floatfix,nofootinbib,
    superscriptaddress
]{revtex4-2}

\usepackage{subfigure}
\usepackage[utf8]{inputenc}
\usepackage{times}
\usepackage[breaklinks]{hyperref}
\usepackage{color}
\usepackage{siunitx}

\usepackage{booktabs}
\usepackage[version=4]{mhchem}
\usepackage{booktabs}

\usepackage{amssymb} 
\usepackage{amsmath}
\usepackage{amsthm}
\usepackage{braket}
\usepackage{mathtools}

\newcommand{\pgi}{Peter Gr\"unberg Institut and Institute for Advanced Simulation,
Forschungszentrum J\"ulich and JARA, 52425 J\"ulich, Germany}

\newcommand{\aachen}{Department of Physics, RWTH Aachen University, 52056 Aachen, Germany}

\newcommand{\mainz}{Institute of Physics, Johannes Gutenberg University Mainz, 55099 Mainz, Germany}

\DeclareSIUnit{\au}{{\,a.u.}}
\renewcommand{\vec}[1]{\mathbf{#1}} 

\newcommand{\hatn}{\hat{\mathbf{n}}}

\newcommand{\abs}{\mathrm{\abs}}
\newcommand{\soi}{\alpha_{\mathrm{R}}}

\usepackage{tikz}
\usepackage{tikz-cd}
\usetikzlibrary{calc}
\usetikzlibrary{decorations.markings}
\usepackage{xcolor}
\usepackage{graphicx}

\makeatletter
\renewcommand\@biblabel[1]{#1.}
\makeatother

\begin{document}

\setcounter{secnumdepth}{2} 

%----------------------------------------------------
% Document title
%----------------------------------------------------

\title{The chiral Hall effect in canted ferromagnets and antiferromagnets}

\author{Jonathan Kipp}
\thanks{These authors contributed equally to this work.}
    \affiliation{\pgi}
    \affiliation{\aachen}
\author{Kartik Samanta}
\thanks{These authors contributed equally to this work.}
    \affiliation{\pgi}

\author{Fabian R. Lux}
\thanks{These authors contributed equally to this work.}
    \affiliation{\pgi}
    \affiliation{\aachen}

\author{Maximilian Merte}

    \affiliation{\pgi}
    \affiliation{\aachen}
    \affiliation{\mainz}
\author{Dongwook Go}
\affiliation{\mainz}
    \affiliation{\pgi}

\author{Jan-Philipp Hanke}

    \affiliation{\pgi}

\author{Matthias Redies}

    \affiliation{\pgi}
    \affiliation{\aachen}
\author{Frank Freimuth}

    \affiliation{\pgi}
    
\author{Stefan Bl\"ugel}

    \affiliation{\pgi}
    
\author{Marjana Le\v{z}ai\'c}

    \affiliation{\pgi}
    
\author{Yuriy Mokrousov}

    \affiliation{\pgi}
    \affiliation{\mainz}

%%%%%%%%%%%%%%%%% END OF PREAMBLE %%%%%%%%%%%%%%%%

\begin{abstract}
The anomalous Hall effect has been indispensable in our understanding of numerous magnetic phenomena. This concerns both ferromagnetic materials, as well as diverse classes of antiferromagnets, where in addition to the anomalous and recently discovered crystal Hall effect, the topological Hall effect in non-coplanar antiferromagnets has been a subject of intensive research in the past decades. Here, we uncover a distinct flavor of the Hall effect emerging in generic canted spin systems.
We demonstrate that upon canting, the anomalous Hall effect acquires a contribution which is sensitive to the sense of imprinted vector chirality among spins. We explore the origins and basic properties of corresponding chiral Hall effect, and closely tie it to the symmetry properties of the system. Our findings suggest that the chiral Hall effect and corresponding chiral magneto-optical effects emerge as useful  tools in characterizing an interplay of structure and chirality in complex magnets, as well as in tracking their chiral dynamics and fluctuations.
\end{abstract}

\maketitle

%----------------------------------------------------
% Document body
%----------------------------------------------------
\date{\today}

\section*{Introduction}
In the past two decades the anomalous Hall effect (AHE) $-$ one of the oldest known manifestations of magnetism in solids $-$ has acquired a major role in testing various new paradigms and phenomena in condensed matter physics~\cite{RMP-AHE}. These include, but are not limited to, the issues related to generation and manipulation of spin currents~\cite{RMP-SHE}, current-induced torques on the magnetization~\cite{Miron,Wadley,RMP-SOT}, electrical detection of topological phases of matter~\cite{Chang2013}, and the emergence of non-collinear spin states~\cite{MacDonald_noncol_2014}. 
While originally explored in ferromagnetic (FM) materials, the AHE has come to occupy a special place in the realm of antiferromagnets (AFMs) as well~\cite{PhysRevLett.87.116801,Libor2018}. While it is well-known that in non-coplanar AFMs the AHE
can arise even without spin-orbit interaction, the AHE emerging in collinear AFMs has been recently discovered~\cite{Libor2020,Libor2020-2}, where the latter crystal Hall effect originates in the breaking of symmetry brought by the non-magnetic cage of atoms via structural chirality~\cite{Libor2020,Kartik,Tsymbal2020}.

The direct relation of the AHE to the geometry and topology of electronic states lends a way to utilizing the AHE as a probe for emergence of various Berry phase properties, which has become one of the major areas of research in the past years. Here, the AHE is traditionally associated with the reciprocal $k$-space Berry phase of Bloch electrons~\cite{Niu_Berry_2010}, while its relation to the real-space Berry phases of electrons in winding spin structures is reflected in celebrated topological Hall effect of systems which exhibit non-vanishing scalar spin chirality $\mathbf{S}_i\cdot(\mathbf{S}_j\times \mathbf{S}_k)$ among neighboring triplets of spins, such as skyrmions~\cite{Fabian-2020}. Recently it has been shown that the $k$-space and real-space Berry phases are closely linked together in giving rise to the so-called chiral Hall effect of spin textures~\cite{Fabian-2020}. In contrast to the AHE in ferromagnets and topological Hall effect of skyrmions, the chiral Hall effect is sensitive to the sense of smooth rotation, or, chirality, of the magnetization in e.g. chiral domain walls~\cite{Fabian-2020}. On the other hand, recent studies show that the effect of spin canting on the electronic structure and the AHE in collinear antiferromagnets can be significant~\cite{Suzuki_2016,Takahashi_2018,HuaChen,Yang_2020}.     

In this work we  demonstrate the emergence of a  distinct flavor of the AHE, which can be prominent both in ferromagnets and antiferromagnets. We show that it arises in 
diverse magnetic systems upon imprinting the vector chirality $\mathbf{S}_i\times\mathbf{S}_j$ among pairs of neighboring spins by canting driven by external fields or thermal fluctuations. We demonstrate that, similarly to its twin in the world of smooth textures, the  chiral Hall effect is sensitive to the sense of vector chirality exhibited by pairs of frustrated spins.
We theoretically investigate the properties of this phenomenon, show that it can be significant in diverse classes of materials, and demonstrate its clear distinction from the conventional anomalous and topological Hall effects by showing that
it has a profoundly different Berry phase origin. 
Importantly, we argue that the inclusion of chiral Hall effect into the palette of complex phenomena exhibited by ferromagnets and antiferromagnets is indispensable for providing a unified categorization of the Hall effects  $-$ which is a prerogative for a conclusive read-out of crystal structure, magnetic order, and dynamics exhibited by complex magnets.
\begin{figure*}[t!]
		\includegraphics[width=0.75\hsize]{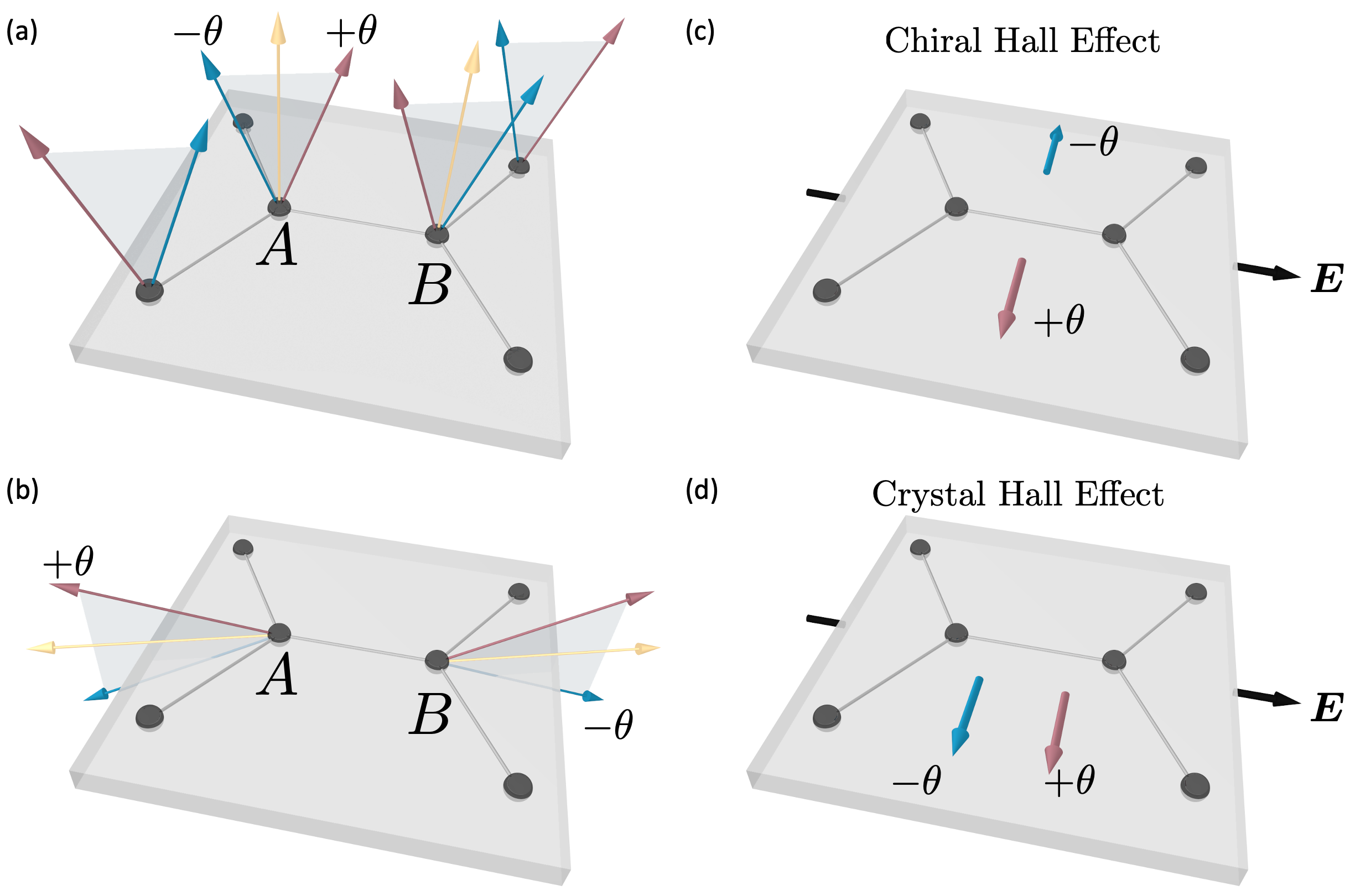}
		\caption{\label{FIG1} {\bf Sketch of the definition of crystal Hall  and chiral Hall  effects in canted ferromangets and antiferromagnets}. Once collinear ferromagnetic or antiferromagnetic order (light yellow arrows in (a) and (b)) is broken by canting with positive ($+\theta$, red arrows) or negative ($-\theta$, blue arrows) sense of vector chirality, the modifications in the electronic structure result in the modifications of the anomalous Hall conductivity (AHC), $\sigma_{xy}(\theta)$. The AHC can decomposed into the crystal Hall (symmetric, $\theta$-even) part, $\sigma_{xy}^s=\left(\sigma_{xy}(+\theta)+\sigma_{xy}(-\theta)\right)/2$, (c), and the chiral Hall (anti-symmetric, $\theta$-odd) part $\sigma_{xy}^a=\left(\sigma_{xy}(+\theta)-\sigma_{xy}(-\theta)\right)/2$, (d). In (c) and (d) the red and blue arrows correspond to the direction of the Hall current for positive and negative chirality in an applied electric field $\mathbf{E}$.}
\end{figure*}

\noindent
%{\large{\bf Results}}\\
\section*{Results}
In this work, we consider the effect of finite vector chirality on the AHE of initially collinear ferro- and antiferromagnetic two-dimensional (2D) systems, which is induced by small canting away from the initial configuration of spins, see (Fig. 1). We concentrate specifically on the case of crystals which comprise two spins in the unit cell, such as a honeycomb lattice
of magnetic atoms, and discuss how our findings can be generalized to the case of several magnetic atom types. Given the original collinear arrangement of spins on sites A and B, $\mathbf{s}_{\rm A}$ and $\mathbf{s}_{\rm B}$, along a certain axis $\hat{\mathbf{s}}_0$, we define a plane which contains this axis as well as spins canted with respect to $\hat{\mathbf{s}}_0$ by an angle $+\theta$ (for $\mathbf{s}_{\rm A}$) and $-\theta$ (for $\mathbf{s}_{\rm B}$).
With this definition, the reversal of sign in the canting angle $\theta\rightarrow -\theta$ provides a state of opposite chirality $\boldsymbol{\chi}$, which we define as $\boldsymbol{\chi}=\mathbf{s}_{\rm A}\times\mathbf{s}_{\rm B}$, with $\chi=|\sin\theta|$, where we assume that the length of the spins does not change upon canting, see (Fig. 1 a,b). In the presence of spin-orbit interaction (SOI) and upon breaking of certain crystalline symmetries, such as inversion symmetry, which is naturally broken upon depositing the 2D magnetic lattice on a surface, the electronic structure of the system with positive chirality 
can be different from that with negative chirality. 

The canting-driven modifications in the electronic structure inevitably result in the modifications brought to the AHE of the system. This aspect presents the focus of our work. In the case of a 2D system considered here, only the $xy$-component of the conductivity tensor which we denote as $\sigma_{xy}$ encodes the information about the magnitude of the AHE. We consider only the intrinsic part of the AHE as given by the $\mathbf{k}$-dependent Berry curvature of the occupied states $\Omega_{xy}(\mathbf{k}) =\sum_{n\in {\rm occ}}2\Im \Braket{\partial_{k_x}u_{n\mathbf{k}}|\partial_{k_y}u_{n\mathbf{k}}} $ where the sum runs over occupied states at point $\mathbf{k}$ and $u_{n\mathbf{k}}$ is the  lattice-periodic Bloch state $n$. The AHC is given by the Brillouin zone (BZ) integral $\sigma_{xy}=\int_{\rm BZ} \Omega_{xy}(\mathbf{k})\,d\mathbf{k}$ 
(see more details in the section Methods). 
In order to track the changes in $\sigma_{xy}$ with respect to canting as given by the angle $\theta$, 
we introduce two key quantities $-$ the symmetric ($\sigma_{xy}^s$) and antisymmetric ($\sigma_{xy}^a$) parts of the anomalous Hall conductivity (AHC) $-$ defined as follows:
\begin{equation}\label{Eq1}
\sigma_{xy}^{s(a)}(\theta) = 
\frac{\sigma_{xy}(\theta) \pm \sigma_{xy}(-\theta)}{2} = \int_{\rm BZ} \Omega_{xy}^{s(a)}(\theta,\mathbf{k})\,d\mathbf{k},
\end{equation}
where the symmetric and antisymmetric parts of the Berry curvature are determined at each $\mathbf{k}$-point as $\Omega^{s(a)}_{xy}(\theta,\mathbf{k})=\left[ \Omega_{xy}(\theta,\mathbf{k}) \pm \Omega_{xy}(-\theta,\mathbf{k}) \right]/2$. The latter dependence of $\Omega_{xy}$  on $\theta$ arises in response to the dependence of electonic states, whose geometry the Berry curvature measures, on canting.

According to its definition, the symmetric AHC  has the same value for the states of opposite chirality,~i.e.~it is $\theta$-even: $\sigma_{xy}^{s}(\theta)=\sigma_{xy}^{s}(-\theta)$, see (Fig. 1 d). Since at zero canting the symmetric AHC is given by the AHC of the collinear system,  $\sigma_{xy}^{s}(\theta=0)=\sigma_{xy}(\theta=0)=\sigma_{xy}^0$, we will refer to this part of the AHC as the crystal
Hall conductivity, as for collinear AFMs it would correspond to the situation of crystal Hall effect~\cite{Libor2019}.
In collinear FMs this would correspond to the conventional definition of the ``ferromagnetic" AHE.
On the other hand, the antisymmetric AHC changes sign when $\theta\rightarrow -\theta$,~i.e.~it is $\theta$-odd: $\sigma_{xy}^{a}(\theta)=-\sigma_{xy}^a(-\theta)$, see (Fig. 1 c), and it vanishes for the collinear configuration. Since this part of the AHC is sensitive to the sense of chirality $\boldsymbol{\chi}$, we refer to it as the chiral Hall conductivity.
This name is further motivated by the fact that the chirality-sensitive Hall effect has been recently discovered in systems where a finite chirality is imprinted by smooth spiral-like deformations of the spin texture~\cite{Fabian-2020}. The chiral Hall effect discussed here presents a version of the latter phenomenon where a specific sense of chirality is generated by lattice-periodic short-wavelength deformations of the spin structure.

By definition, both effects $-$ the crystal Hall and chiral Hall  effects $-$ when added together, provide the total AHC of the system: $\sigma_{xy}^{s}(\theta) + \sigma_{xy}^{a}(\theta)=\sigma_{xy}(\theta)$.
However, while the crystal Hall effect picks up even powers of $\theta$ in the Taylor expansion of $\sigma_{xy}(\theta)$ around the collinear state, $\sigma_{xy}^{s}(\theta)=\sigma_{xy}^0 + a\theta^2 + ...$, the chiral Hall effect accumulates odd terms in the latter expansion, $\sigma_{xy}^{a}(\theta)=b\theta +c\theta^3 + ...$, where coefficients $a,b$ and $c$ depend on the electronic structure in the collinear state. This tells us, that in the limit of small canting (i.e.~to the first order in $\theta$) the deviations of $\sigma_{xy}$ from $\sigma_{xy}^0$ are manifestly chiral in nature. Correspondingly, understanding the properties of the chiral Hall effect is of utter importance for understanding the behavior of the AHE in collinear magnets where the spins are canted either as a result of external electric and magnetic fields, chemical or structural tuning of exchange interactions, and thermal fluctuations. 

\begin{figure*}[ht!]
	\includegraphics[width=0.90\hsize]{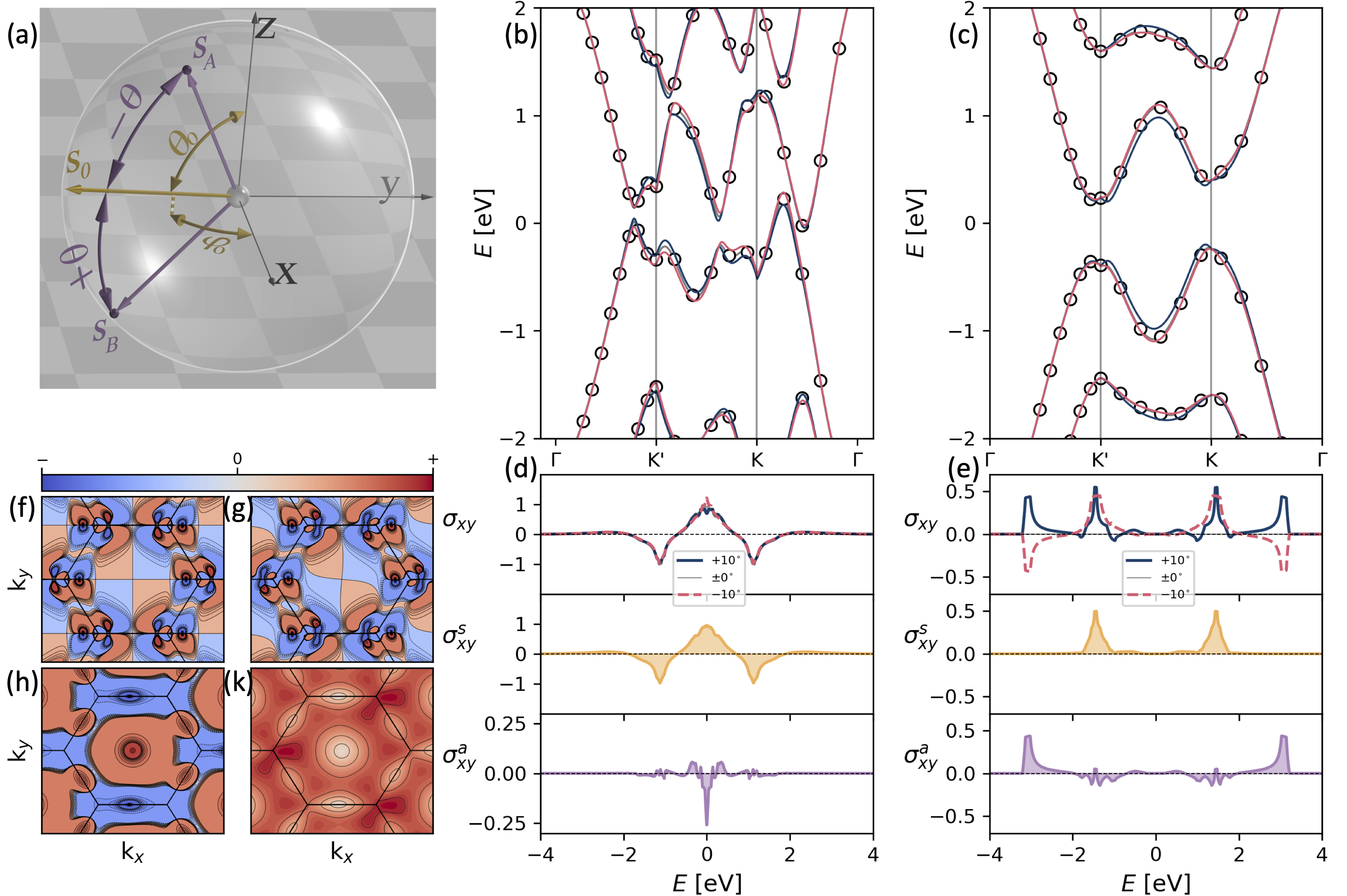}
	\caption{\label{FIG2} {\bf  The emergence of chiral and crystal Hall effect of ferro- and antiferromagnets on a honeycomb lattice}.
		(a) The definition of the angles used to characterize the canted spin structure of spins $\mathbf{s}_{\rm A}$ and $\mathbf{s}_{\rm B}$. The initial direction of collinear magnetization $\hat{\mathbf{s}}_0=(\theta_0,\varphi_0)$ with polar angle $\theta_0$ and azimuthal angle $\varphi_0$ is kept constant during canting, $\hat{\mathbf{s}}_0\sim\mathbf{s}_{\rm A}+\mathbf{s}_{\rm B}$. The spins are canted in the plane of constant $\varphi_0$ by an angle $\theta$ for $\mathbf{s}_{\rm A}$ and $-\theta$ for $\mathbf{s}_{\rm B}$ with respect to $\hat{\mathbf{s}}_0$. (b-c) The changes in the bandstructure of the ferromagnetic (FM) (b) and antiferromagnetic (AFM) (c) spins initially along $\hat{\mathbf{s}}_0=(100^{\circ},10^{\circ})$ upon canting by $\pm 10^{\circ}$. The thin grey line with circles marks the initial bandstucture while blue and red lines mark the bandstructure for $\theta= 10^{\circ}$ and $\theta=-10^{\circ}$, respectively. (d-e) The corresponding anomalous Hall conductivity (AHC), $\sigma_{xy}$, as a function of the Fermi energy is shown for the FM (d) and AFM (e) cases for positive (solid blue line) and negative (dashed red line) canting. The symmetric, $\sigma_{xy}^s$, and anti-symmetric, $\sigma_{xy}^a$, parts of the AHC are shown with dark orange and dark blue lines. All values are in  $e^2/h$, where $e$ is the elementary charge and $h$ is Planck's constant.
		(f-k) While for the high-symmetry direction of $\hat{\mathbf{s}}_0=(100^{\circ},0^{\circ})$ the symmetry properties of the Berry curvature of the first two bands in the FM case, $\Omega^a(10^{\circ},\mathbf{k})$, lead to vanishing overall chiral Hall effect, (f), the breaking of symmetry for $\hat{\mathbf{s}}_0=(100^{\circ},10^{\circ})$ results in a net effect, (g). The complex structure of $\Omega^a(10^{\circ},\mathbf{k})$ of the first band from (c) in $\mathbf{k}$-space, (h), is clearly correlated with the separation between the first and second bands in energy, shown in (k). }
\end{figure*}

\noindent
{\bf Model considerations.} We start by considering the existence and properties of the chiral Hall effect on a bi-partite honeycomb lattice of magnetic spins. The effective lattice tight-binding Hamitonian reads:
\begin{equation}
\begin{split}
H = -t \sum\limits_{\langle ij \rangle\alpha}  c_{i\alpha}^\dagger c_{j\alpha}^{\phantom{\dagger}} &+ i \alpha_{\rm R}\sum\limits_{\langle ij \rangle\alpha \beta}  \hat{\mathbf e}_z \cdot (\boldsymbol{\sigma} \times {\mathbf d}_{ij})_{\alpha\beta}\, c_{i\alpha}^\dagger c_{j\beta}^{\phantom{\dagger}}\\
&+ \lambda_{\rm ex} \sum_{i\alpha \beta} (\hat{\mathbf s}_i\cdot \boldsymbol{\sigma})_{\alpha\beta}\, c_{i\alpha}^\dagger c_{i\beta}^{\phantom{\dagger}},
\end{split}
\label{eq:model}
\end{equation}
where $c_{i\alpha}^\dagger$ ($c_{i\alpha}^{\vphantom{\dagger}}$) denotes the creation (annihilation) of an electron with spin $\alpha$ at site $i$, $\langle ...\rangle$ restricts the sums to nearest neighbors, the unit vector $\mathbf d_{ij}$ points from $j$ to $i$, and $\boldsymbol{\sigma}$ stands for the vector of Pauli matrices. Besides the hopping with amplitude $t$,  Eq.~\eqref{eq:model} contains the Rashba spin-orbit coupling of strength $\alpha_\text{R}$ originating for example in the surface potential gradient. The remaining term in equation~\eqref{eq:model} is the local exchange term with $\lambda_{\rm ex}$ characterizing the strength of exchange splitting and $\hat{\mathbf{s}}_i$ stands for the direction of spin on site $i$. 

Here, we work with the following parameters of the model: $t=1.0$\,eV, $\alpha_{\rm R}=0.4$\,eV, and $\lambda_{\rm ex}=1.4$\,eV. We start with the initial direction of atomic spins along a given direction $\hat{\mathbf{s}}_0$ characterized with polar angles $\hat{\mathbf{s}}_0=(\theta_0,\varphi_0)$, see (Fig. 2 a), with $\hat{\mathbf{s}}_{\rm A}$ and $\hat{\mathbf{s}}_{\rm B}$ along $\hat{\mathbf{s}}_0$ for a FM, and with $\hat{\mathbf{s}}_{\rm A}=-\hat{\mathbf{s}}_{\rm B}=\hat{\mathbf{s}}_0$ in case of an AFM configuration. 
Following the symmetry analysis (see Supplementary Note 1), we consider the canting plane which is orthogonal to the $xy$-plane and which contains $\hat{\mathbf{s}}_0$. Within this plane, the azimuthal angle of all spins is constant and the canting is characterized by an angle $\pm\theta$ away from $\hat{\mathbf{s}}_0$ for $\hat{\mathbf{s}}_{\rm A/B}$. A change of sign of $\theta$ corresponds to switching the sign of the chirality among $\hat{\mathbf{s}}_{\rm A}$ and $\hat{\mathbf{s}}_{\rm B}$, (Fig. 2 a). 

Before proceeding with the analysis of the AHE, we inspect the influence of chirality on the band structure of the model. To do this, we choose the initial collinear direction of the spins along $\hat{\mathbf{s}}_0=(100^{\circ},10^{\circ})$, which breaks all symmetries in the system. The bandstructures of the FM  and AFM configurations for the collinear as well as canted by $\pm 10^{\circ}$ cases are shown in 
(Fig. 2 b) and (c), respectively. The band structure for the FM case for $\hat{\mathbf{s}}_0=(90^{\circ},0^{\circ})$ is known to be gapped at half-filling, where the gap of the system is topologically non-trivial~\cite{Niu_mixed_2019}. Clearly, canting-driven band dynamics is different for two opposite chiralities, and respective band shifts sensitively depend on the structural properties. They can be further separated into contributions which are even and odd in the Rashba strength. Among these, the ones odd in $\alpha_{\rm R}$,~i.e.,~sensitive to the sense of structural chirality, are closely related to the emergence of  Dzyaloshinskii-Moriya interaction among spins $\mathbf{s}_{\rm A}$ and  $\mathbf{s}_{\rm B}$~\cite{Dzyalosinskij_1957,Moriya_1960,Bode}.

In the FM case, the chiral band shifts observed in (Fig. 2) are directly related to the sense of inversion symmetry breaking via the Rashba term in Eq.~\ref{eq:model} and corresponding structural chirality: upon changing the sign of $\alpha_{\rm R}\rightarrow -\alpha_{\rm R}$ in the Hamiltonian, the bands of the configurations with opposite chirality simply exchange their energetic position. 
The latter effect can be also understood based on an effective gauge theory, applied recently to the study of orbital magnetism in chiral spin systems~\cite{Lux2018}, where the effect of canting and generally vector chirality was shown to be equivalent to an effect of a fictitious chiral magnetic field $B^{\rm eff}_{\rm R}\sim\boldsymbol{\chi}$, applied to a collinear FM system. Within the interfacial Rashba model it can be shown analytically that $B^{\rm eff}_{\rm R}\sim \alpha_{\rm R}$, implying that  $B^{\rm eff}_{\rm R}$ changes sign when the sense of inversion symmetry breaking is reversed. Consequently, the corresponding band shifts of the ferromagnetic electronic states of Hamiltonian~(\ref{eq:model}), a lattice realization of the interfacial Rashba model, change sign.

In ferromagnets with broken inversion symmetry the emergence of non-vanishing chiral magnetic field generated by chiral spin canting goes hand in hand with the rise of the linear-in-chirality contribution to the Hall effect $-$ the chiral Hall effect. 
Our analysis clearly reveals that the chiral Hall effect is a general effect appearing not only in smooth textures~\cite{Fabian-2020} but also in the context of canted FMs. In (Fig. 2 d) we show explicit calculations of $\sigma_{xy}$ (for $+\theta$ and $-\theta$ with $\theta=10^{\circ}$), $\sigma_{xy}^s$ and $\sigma_{xy}^a$ for $\hat{\mathbf{s}}_0=(100^{\circ},10^{\circ})$ as a function of band filling of the model.
We observe that  significant dependence of the band structure on the chirality results in a noticeable influence of chirality on the AHC mainly  close to half-filling. The symmetric in chirality $\sigma_{xy}^s$ largely follows the behaviour of $\sigma_{xy}^0$ in the whole range of energies, while the behavior of the $\sigma_{xy}^a$ is correlated with fine canting-driven band dynamics reflected in a complex distribution of the anti-symmetric Berry curvature in $k$-space, shown in (Fig. 2 g) for the lowest two bands. 
And while the latter distribution does not vanish $k$-point-wise for any direction of $\mathbf{s}_0$ except for the case when $\theta=n\pi, n\in\mathbb{Z}$, the overall BZ integral of the antisymmetric Berry curvature vanishes owing to mirror symmetry for high-symmetry directions of $\mathbf{s}_0$ with $\varphi=n\pi/3$, see~e.g.~(Fig. 2 f).

The pronounced chiral Hall effect of the FM model at half-filling is closely related to the topological phase transition occuring for $\hat{\mathbf{s}}_0=(90^{\circ},0^{\circ})$. Here, as the direction of the collinear magnetization passes through $(xy)$-plane, the quantized Hall conductance of the system changes by $2\frac{e^2}{h}$ in response to the change in the chirality of the Chern insulating state. This topological phase transition is the consequence of the presence of a so-called mixed Weyl point in the electronic structure at $E_F=0$\,eV for the in-plane magnetization~\cite{Hanke_mixed_2017-2}, the Berry phase nature of which we discuss later. Correspondingly, energy-resolved calculations of the chiral Hall conductivity as a function of the angle $\theta_0$, presented in (Fig. 3 a), reveal a pronounced and very complex structure of $\sigma_{xy}^a$ next to the mixed Weyl point, which stands in contrast to a relatively smooth behavior of $\sigma_{xy}^s$ in $(\theta_0,E_F)$-space (not shown). On the other hand, the chiral Hall effect exhibits a much stronger response to the canting angle $\theta$, as compared to $\sigma_{xy}^s$: as shown in (Fig. 3 b) for the case of half-filling, while $\sigma_{xy}^s$ changes by about 0.05\,$e^2/h$ for the canting angle of up to $10^{\circ}$, in the same range of $\theta$ the corresponding change of $\sigma_{xy}^a$ is larger by an order of magnitude. In accordance to arguments from above, the general trend of $\sigma_{xy}^a$ and $\sigma_{xy}^s$ with $\theta$ is linear and quadratic, respectively, when the canting angle is sufficiently small.   

In contrast to a ferromagnet, for the antiferromagnetic case the magnitudes of the crystal and chiral Hall effects are large and comparable, but they manifest in different energy regions, see (Fig. 2 e). The AFM case presents another example of a correlation between the antisymmetric Berry curvature and the electronic structure: as visible in (Fig. 2 h,k) the emergence of strong features in the Berry curvature of the first band of the model is consistent with the first and second band coming close to each other in energy at specific points in the BZ.
In analogy to ferromagnets, this gives rise to monopoles of  special type which manifest in an enhanced antisymmetric Berry curvature, as discussed below.
In analogy to the FM case considered above, the scaling of the chiral Hall effect with the canting angle can be confirmed to be linear for small $\theta$, see~e.g.~the inset of (Fig. 3 b).

Overall, as we have shown above by explicit calculations, the flavor of the Hall effect linear in spin chirality $-$ the chiral Hall effect $-$ exists and can be prominent both in FMs and AFMs. In the next two sections we uncover the nature of the chiral Hall effect as a phenomenon which can be clearly distinguished from the ``conventional" AHE, associated with the change in the overall magnetization of the system. For FMs, the conceptual difference between the two is very clear, as both of the canted states, used to arrive at the chiral Hall effect, (Fig. 3 a), share the same overall magnetization. How to draw the distinction for AFMs is less obvious, as the change in chirality in (Fig. 3 b) is associated with the change in sign and magnitude of the overall ``ferromagnetic" magnetization arising upon canting. Below, we formalize the classification of chiral and crystal Hall effects consistently in canted ferro- and antiferromagnets, referring to symmetry arguments. 

\noindent
{\bf Symmetry analysis.}
The magnetic order is fully characterized by the staggered field $\vec{n}_-$ and the ferromagnetic field $\vec{n}_+$ which are defined according to $\vec{n}_\pm = \vec{s}_\mathrm{A} \pm \vec{s}_\mathrm{B} $.
The Hall conductivity can thus be decomposed into terms which are even and odd with respect to the interchange of $\hatn_- \to - \hatn_-$, i.e.,
\begin{align}
\sigma_{xy}  (\vec{n}_+ ;\vec{n}_- )
& = 
\sigma_{xy}^\mathrm{odd}  (\vec{n}_+ ;\vec{n}_- )
+
\sigma_{xy}^\mathrm{even}  (\vec{n}_+ ;\vec{n}_- ).
\end{align}
The off-diagonal components of the conductivity as they arise from the Berry curvature can be interpreted as the components of an axial vector which is odd under time-reversal.
Each of these terms can thus be further expanded as a sum over all terms which are odd under magnetization reversal:
\begin{align}
\sigma_{xy}^\mathrm{odd} &= \sum_{k,l=0}^\infty (c^\mathrm{odd}_{xy})^i   : ( \vec{n}_-^{\otimes 2k+1}   \otimes  \vec{n}_+^{\otimes 2l} )_i
\label{eq:sigma_expansion_A}
\\
\sigma_{xy}^\mathrm{even} &= \sum_{k,l=0}^\infty (c^\mathrm{even}_{xy})^i :  ( \vec{n}_-^{\otimes 2k}   \otimes  \vec{n}_+^{\otimes 2l+1} )_i,
\label{eq:sigma_expansion_B}
\end{align}
where $:$ denotes the tensor contraction over the multi-index $i=(i_1, \ldots, i_{2(k+l)+1})$ (we refer to the Supplemental Note 1 for an explicit example).
This decomposition into odd and even parts also corresponds to the parity under magnetic sublattice interchange, which would leave $\hatn_+$ invariant.
Therefore, the symmetry requirements for these two tensors are quite different.
In order for $\sigma_{xy}^\mathrm{even}$ to be finite, the crystal symmetry needs to support axial tensors of odd order. 
\begin{figure}[t!]
	\begin{center}
		\rotatebox{0}{\includegraphics[width=0.45\textwidth]{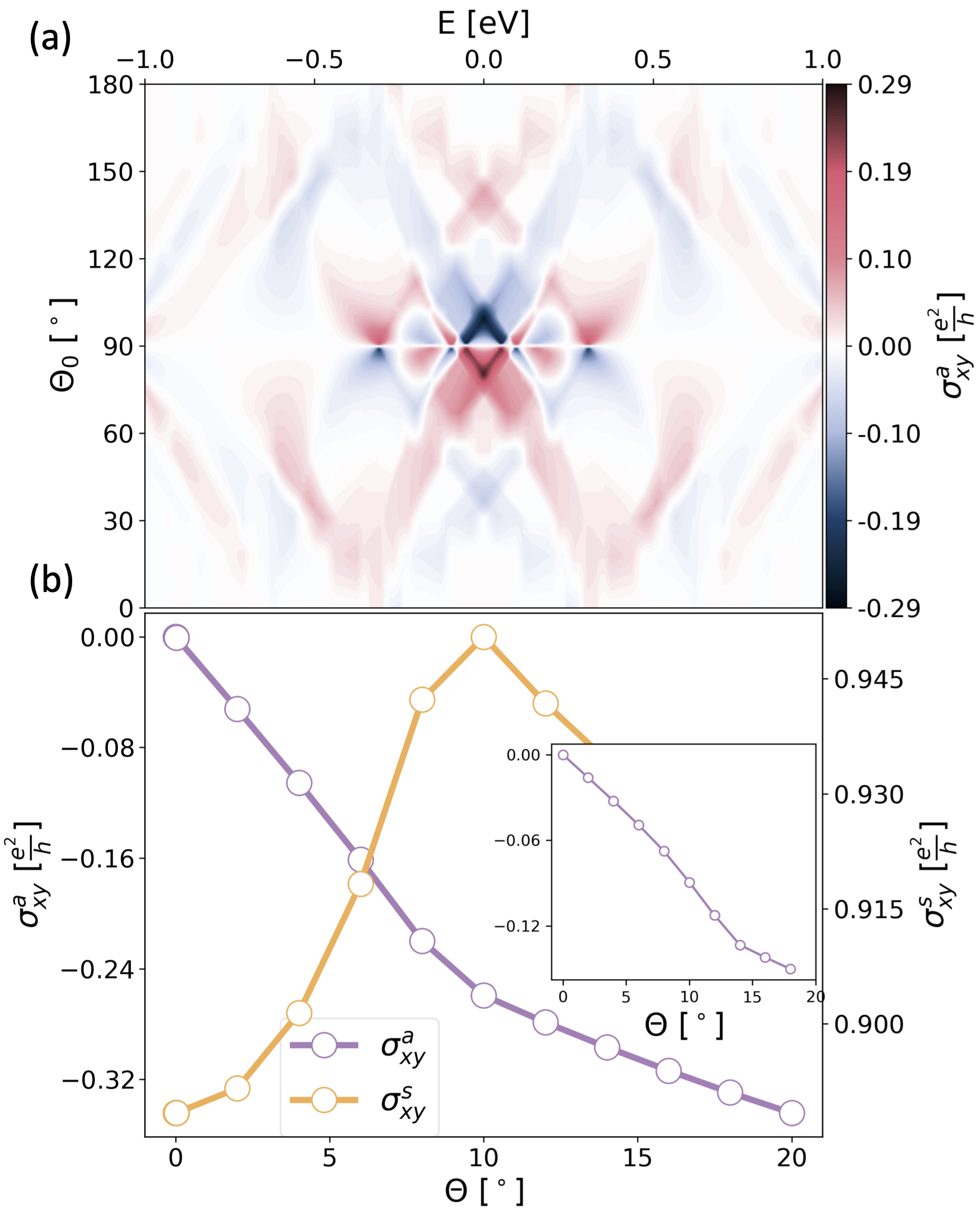}}
	\end{center}
	\caption{{\bf Properties of the chiral Hall effect}. (a) Behavior of the antisymmetric part of the  anomalous Hall conductivity $\sigma_{xy}^a$ at $10^{\circ}$ canting as a function of Fermi energy and direction of collinear ferromagnetic magnetization $\mathbf{s}_0=(\theta_0,10^{\circ})$. While the fine structure of the chiral Hall effect correlates with the band structure dynamics in response to canting and rotation of the initial magnetization, the origin of the effect in the Weyl point at half filling for $\theta_0=90^{\circ}$, serving as  a source of staggered mixed Berry curvature, is visible. (b) The scaling of the crystal (orange line) and chiral (violet line) Hall effects with the canting angle $\theta$ at half-filling of the ferromagnetic case $\mathbf{s}_0=(100^{\circ},10^{\circ})$. The inset displays the scaling of the chiral Hall effect with $\theta$ for Fermi energy $E_F=-1.5$\,eV in the antiferromagnetic case with the same $\mathbf{s}_0$. 
	}
	\label{FIG3}
\end{figure}

In particular, the effect is then even under lattice inversion and in our model it is thus necessarily even in the spin-orbit coupling strength $\soi$.
The case is different for $\sigma_{xy}^\mathrm{odd}$, whose tensorial components above either transform axial or polar depending on whether or not the symmetry under consideration interchanges the lattice sites: since $P \vec{s}_\mathrm{A/B} =  \vec{s}_\mathrm{B/A}$ for the inversion operation $P$, the staggered magnetization would behave polar for our lattice, i.e.,  $P \vec{n}_- = - \vec{n}_-$, and not axial as $\vec{n}_+$.
For small values of the spin-orbit strength, $\sigma_{xy}^\mathrm{odd}$ is therefore linear in $\soi$ (generally odd in $\soi$), which is a corollary to the general fact that polar tensors of odd rank are identically zero in centrosymmetric crystal structures, see Table~I.

While the general expansion in Eqs.~(\ref{eq:sigma_expansion_A}-\ref{eq:sigma_expansion_B}) is in principle complete,
a formulation in terms of the chirality $\boldsymbol{\chi}$ offers a deeper insight into the various effects which can appear in ferro- and antiferromagnets.
Based on the definitions above the chirality itself can be reinterpreted as
\begin{equation}
\boldsymbol{\chi} = \vec{s}_\mathrm{A} \times \vec{s}_\mathrm{B} = \frac{1}{2} (
\vec{n}_- \times \vec{n}_+ ),
\end{equation}
which is therefore odd in both $\vec{n}_-$ and $\vec{n}_+$, but even under time-reversal.
Since $\vec{n}_+ \cdot \vec{n}_- = 0$, one has
$
\boldsymbol{\chi} \times \vec{n}_\pm = \mp \| \vec{n}_\pm \|^2 \vec{n}_\mp /2
$.
Hence, the leading order terms in the expansion of $ \sigma_{xy}^\mathrm{odd}$
and $ \sigma_{xy}^\mathrm{even}$
can be written in two equivalent ways by either replacing all appearing $\vec{n}_-$ or $\vec{n}_+$ factors in terms of chirality, i.e.,
\begin{align}
\sigma_{xy}^\mathrm{odd} &\sim  \sum_i \alpha^\mathrm{FM}_i ( \hatn_+) \chi_i =
\sum_{ij} \alpha^\mathrm{AFM}_{ij} ( \hatn_-) \chi_i \chi_j
\\
\sigma_{xy}^\mathrm{even} &\sim  \sum_i \beta^\mathrm{AFM}_i ( \hatn_-) \chi_i  =
\sum_{ij} \beta^\mathrm{FM}_{ij} ( \hatn_+) \chi_i  \chi_j,
\end{align}
where $\alpha^\mathrm{FM}_i$, $\alpha^\mathrm{AFM}_{ij}$  and $\beta^\mathrm{FM}_i$, $\beta^\mathrm{AFM}_{ij}$ are odd under time-reversal.
The choice of $\alpha$ and $\beta$ coefficients is a matter of philosophy. In a weakly canted ferromagnet, for example, it makes sense to formulate the change in conductivity as response to the $\boldsymbol{\chi}$ where the coefficients depend only on the electronic structure of the unperturbed, collinear system, which is solely determined by $\hatn_+$.
For a weakly canted antiferromagnet, it makes sense to do the opposite.
This situation is summarized in Table \ref{tab:categorization}. 

The chiral Hall effect can be now understood as the effect which accumulates all terms containing an odd number of $\chi_i$ relative to their collinear reference state.
To lowest order, these are therefore linear in $\chi_i$ and hence chiral.
This definition corresponds exactly to the way the chiral Hall effect has been defined at the beginning
and it  corresponds also to the diagonal terms in Table~I. In particular, as we show in the Supplemental Note 1,  the ``topological" terms of the type $\hat{\mathbf{n}}_{+}\cdot\boldsymbol{\chi}$ do not appear in the expansions of the conductivities above explicitly, which allows to draw a strict line between the chiral Hall effect, and the topological Hall effect rooted in the scalar spin chirality. On the other hand, 
 the crystal Hall effect can be identified with those terms which are even in $\chi_i$ when formulated with respect to the collinear reference state.
For the canted antiferromagnet, this corresponds to the definition given in~Ref.~\cite{Libor2020}, which we extend here to the case of canted ferromagnets. 
The lowest order introduced by the canting is thus bichiral, i.e. it is quadratic in $\chi_i$. This corresponds to off-diagonal terms in Table~I, which thus provides  complete characterization of  flavors of the Hall effect in terms of chirality of the spin structure.

Note that the expansion of $\sigma_{xy}^\mathrm{even}$ in Eq.~(\ref{eq:sigma_expansion_B}) also contains the contribution from the usual anomalous Hall effect, which is the lowest order term proportional to the magnetization $\hatn_+$.
The chiral Hall effect in AFMs and the crystal Hall effect in FMs, while being formally proportional to $\hatn_+$, are different from the conventional AHE contribution as their structure is generally more complex, and the corresponding coefficients in  the expansion  (\ref{eq:sigma_expansion_B})  depend on the electronic structure in a different way than the usual AHE coefficient.
This is directly reflected in the different Berry phase nature of the two classes of phenomena.
Below, we provide the geometrical theory of the chiral Hall effect, which marks it as a playground for exploring novel types of Berry phases, not accessible in the realm of AHE of collinear magnets.

\noindent
{\bf Berry phase picture of chiral Hall effect.} We show that the chiral Hall effect allows for an elegant interpretation in geometrical terms which relate the geometry of Bloch electronic states in $k$-space with the geometry associated with spin rotations. To do this, we consider a perturbation of the system which is characterized by a parameter $\lambda(\theta)$ corresponding  to staggered infinitesimal rotation of spins on two sublattices by an angle $\theta$ around a fixed direction, as defined before. This type of perturbation is distinctly different from that associated with a variation of the total magnetization of a collinear FM system, related to the change in the exchange coupling strength, when treated on the model level. 

\begin{table}
	\centering
	\caption{Unified categorization of various Hall effects taking place in canted ferromagnets (FM) and antiferromagnets (AFM) as a function of ferromagnetic/staggered magnetization $\hatn_{+/-}$ and structural chirality $\vec{\chi}$. Here, $\alpha^\mathrm{FM}_i$ and $\beta^\mathrm{FM}_{ij}$ are expansion coefficients, depending on whether the reference state is FM or AFM. The leading order is linear or quadratic in the Rashba spin-orbit interaction parameter $\alpha_R$.}
	\begin{tabular}{cccc}
		\toprule 
		$\vec{s}_\mathrm{A} \leftrightarrow \vec{s}_\mathrm{B}$     & Canted ferromagnet & Canted antiferromagnet  \\ \midrule
		& Chiral Hall Effect & Crystal Hall Effect \\
		$\sigma_{xy}^\mathrm{odd}$ &  $\alpha^\mathrm{FM}_i ( \hatn_+) \chi_i$ & $\alpha^\mathrm{AFM}_{ij} ( \hatn_-) \chi_i \chi_j$ \\
		&  $\sim\alpha_{\rm R}$ & $\sim\alpha_{\rm R}$ \\
		\midrule
		& Crystal Hall Effect &  Chiral Hall Effect     \\
		$\sigma_{xy}^\mathrm{even}$ & $\beta^\mathrm{FM}_{ij} ( \hatn_+) \chi_i \chi_j$ &  $\beta^\mathrm{AFM}_{i} ( \hatn_-) \chi_i $\\
		&  $\sim\alpha_{\rm R}^2$ & $\sim\alpha_{\rm R}^2$ \\
		\bottomrule
	\end{tabular}
	\label{tab:categorization}
\end{table}

We look at the evolution of the $k$-space Berry curvature $\Omega_{xy}$ with $\lambda$, which is ultimately related to the change in the AHC of the system. Namely, we single out the linear in $\lambda$ term by looking at the quantity
$\delta\Omega_{xy} = \lim_{\lambda\rightarrow 0}\partial_\lambda\Omega_{xy}$, which stands for the magnitude of the response of chiral Hall conductivity  to infinitesimal canting,~i.e.~$\Omega_{xy}^a\approx |\theta|\cdot\delta\Omega_{xy}$. Using perturbation theory arguments, it can be shown that at zero temperature (omitting the Fermi surface contribution) 

\begin{equation}\label{mixed}
\delta\Omega_{xy} 
= \Im~\mathrm{tr}_\mathrm{occ}\left(
[\Omega_{xy},\mathcal{A}_{\lambda}
]  +
[
\mathcal{Q}_{\lambda x},\mathcal{A}_{y}
]  +
[\Omega_{y\lambda},\mathcal{A}_{x}
]\right) / 2
,
\end{equation}
antisymmetrized with respect to $( x \leftrightarrow y)$ interchange of indeces, where $\mathcal{A}_{\alpha}=i\Braket{u_n|\partial_{\alpha}u_m}$ with $\alpha=\{k_x,k_y,\lambda\}$ are the components of the Berry connection, $\mathcal{Q}_{\alpha\beta}=\partial_\alpha  \mathcal{A}_\beta + \partial_\beta \mathcal{A}_\alpha$ is the quantity related to the quantum metric tensor~\cite{AA}, and 
$\Omega_{x\lambda}=2\Im\Braket{\partial_{k_x}u_n|\partial_{\lambda}u_m}$ is the mixed component of the Berry curvature tensor. The details on the derivation can be found in Supplemental Note 2.

The appearance in Eq.~(\ref{mixed}) of the mixed Berry curvature, which couples the changes in the electronic states with respect to the Bloch vector to their variation in response to chiral $\theta$-canting, is worth noting. We refer to this type of Berry curvature as the staggered mixed Berry curvature, to distinguish it from the type of the mixed Berry curvature which was introduced in the past for the situation where $\lambda$ 
represents an infinitesimal rotation of the same sense on both atoms, and which corresponds to a coherent rotation of the ferromagnetic or staggered antiferromagnetic magnetization in collinear FMs and AFMs.
The latter type of the Berry curvature was shown to be directly related to the anti-damping spin-orbit torque that an electric field exerts on the collinear magnetization~\cite{Frank-SOT,Hanke_mixed_2017-2,Niu_mixed_2019}. The staggered mixed Berry curvature is thus directly related to the staggered spin-orbit torque, able to drive canting  in collinear systems, which we discuss at a later point.
In fact, Eq.~(\ref{mixed}) is valid for the type of perturbation which corresponds to a coherent rotation as well, which fundamentally relates the spin-orbit torque to the linear in $\theta$ anisotropy of the anomalous Hall conductivity of the collinear system.

\begin{figure}[t!]
	\begin{center}
		\rotatebox{0}{\includegraphics[width=0.45\textwidth]{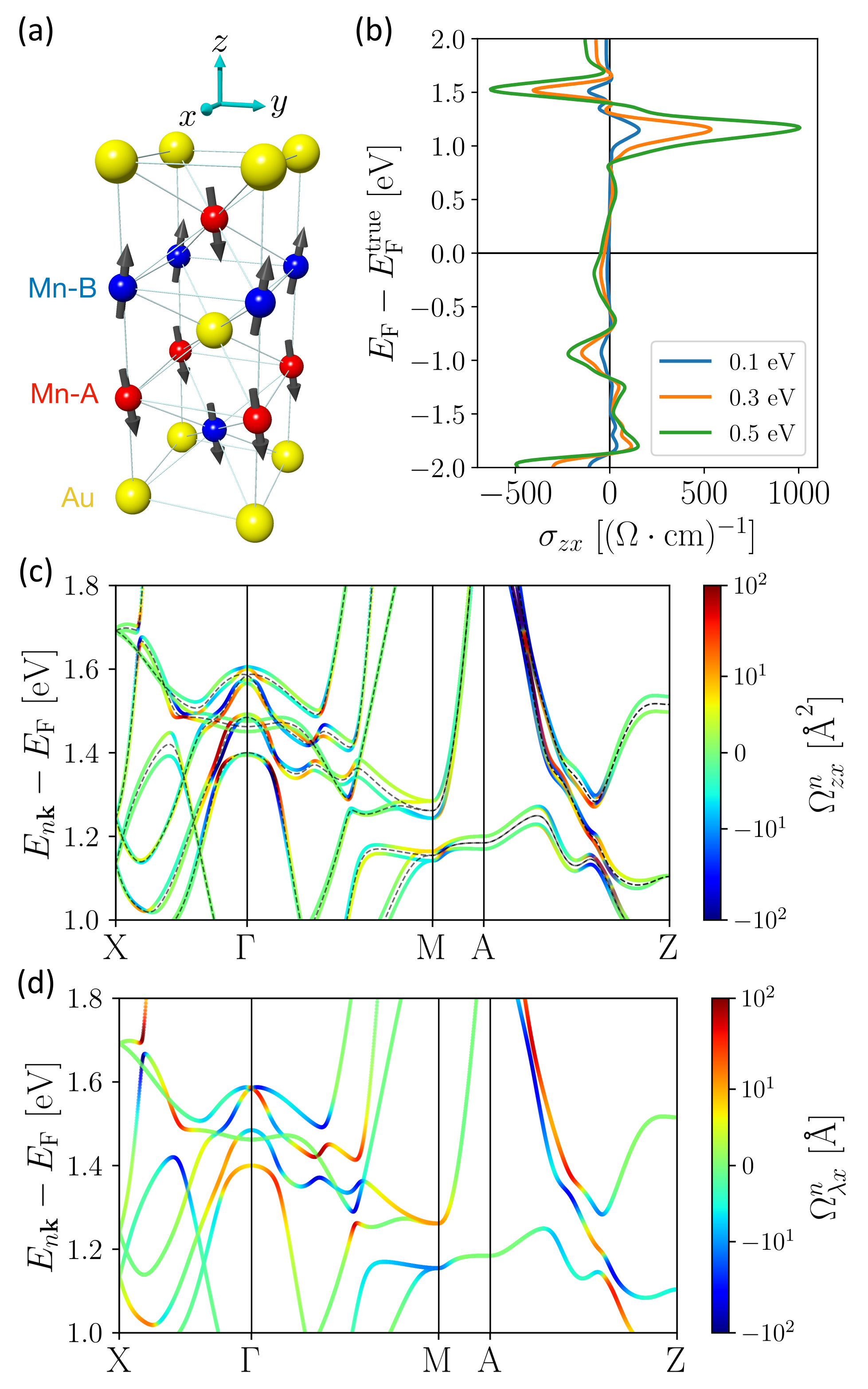}}
	\end{center}
	\caption{{\bf Chiral Hall effect and staggered mixed Berry curvature in Mn$_2$Au.} (a) Crystal structure of Mn$_2$Au with Mn atoms in sublattices A and B denoted with red and blue balls, respectively, and Au atoms shown with yellow balls. The canting of spins, initially oriented along $z$, is induced by applying an exchange field along $y$.  (b) Fermi energy dependence of the chiral Hall conductivity for different strength of the exchange field (100\,meV corresponds to 2$^{\circ}$ canting). (c) Band distribution of $\mathbf{k}$-space Berry curvature $\Omega^n_{zx}$ for electronic states between $+1.0$\,eV and $+2.0$\,eV above the Fermi energy, where the chiral Hall effect is pronounced, for the canting of $+2^{\circ}$. Dashed line indicates the doubly degenerate electronic band structure in the absence of  canting. The effect of opposite canting is identical, with the sign of the Berry curvature of each band reversed. (d) Band distribution of staggered mixed Berry curvature $\Omega^n_{\lambda x}$ for electronic states without canting shown with dashed line in (c). Note that $\Omega^n_{\lambda x}$ is identical for each of the doubly degenerate bands. The correlation between the chiral Hall effect and staggered mixed Berry curvature is evident.}
	\label{}
\end{figure}

The uncovered relation between the 
anomalous Hall effect and
chiral Hall effect with the mixed and 
staggered mixed Berry curvature, respectively, 
is not too surprising. This is easiest understood by referring to the magnetic graphene model studied here. For a collinear case, this model exhibits a band degeneracy of the mixed Weyl type~\cite{Hanke_mixed_2017-2} for the in-plane direction of the magnetization, whose non-zero topological charge is determined by integrating the Berry curvature vector field, constructed out of $k$-space and mixed components of the Berry curvature tensor, around it. The two types of Berry curvature in the vicinity of the mixed Weyl point thus become intertwined with each other by non-trivial topology of the mixed Weyl point. The fundamental relation~(\ref{mixed}) is the formal generalization of this rationale to the situation of a general driving parameter $\lambda$. For our FM model, the pronounced chiral Hall effect in the vicinity of the in-plane magnetization  (Fig. 3 a), which underlines the staggered mixed nature of the band degeneracy, goes hand in hand with large variation of the collinear AHE and large mixed Berry curvature around the degeneracy point, found in the past~\cite{Hanke_mixed_2017-2}. 
The emergence of such staggered mixed Weyl points in the electronic structure correspondingly results in a large response of the AHE to canting, found for instance in~\cite{Suzuki_2016,Takahashi_2018,Yang_2020}, large response in terms for the so-called chiral orbital magnetization~\cite{Lux2018,Fabian-2020}, and a large chiral Hall effect, in accordance to our calculations.

\begin{figure*}[ht!]
	\begin{center}
		\rotatebox{0}{\includegraphics[width=0.93\textwidth]{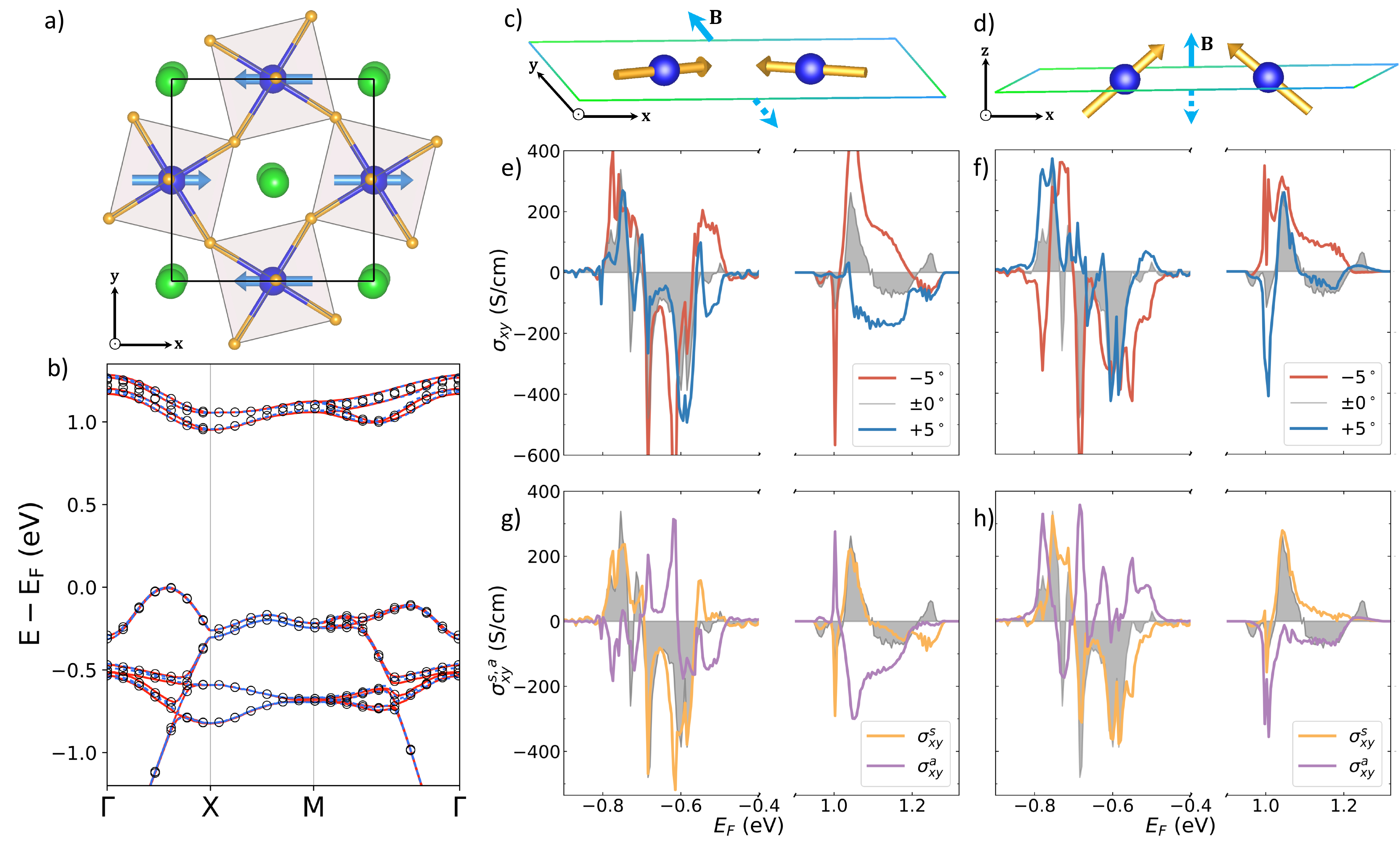}}
	\end{center}
	\caption{{\bf Chiral and crystal Hall effect in monolayer of antiferromagnetic  SrRuO$_3$ (SRO)}.
		(a) Top view of the monolayer with staggered magnetization along $x$. Green, blue and orange  spheres mark Sr, Ru and O atoms, respectively, with arrows representing Ru spins. Visible is the octahedral distortion of oxygen cage surrounding Ru atoms (rotation in the $xy$-plane and tilt with respect to the $z$-axis).
		(b) Band structure of SRO monolayer  with spins along $x$ (black line, open circles), and in the canted state with canting angle of $\theta = \pm 5^{\circ}$ in the $xy$-plane with respect to the $x$-axis (green and red lines for positive and negative chirality, respectively). (c) Schematic of the geometrical setup: Canted state considering the canting angle $\theta = \pm 5^{\circ}$ in the plane of the SRO film ($xy$-plane) with respect to the $x$-axis.
		(d) Same as in (c) for the $xz$-plane of canting along $z$.
		(e-f) Computed anomalous Hall conductivity (AHC) as a function of Fermi level position in the collinear (along $x$) as well as in the canted state. The corresponding geometrical setup is shown schematically in (c) and (d) respectively. Shaded grey areas corresponds to the AHC in the initially collinear state, $\sigma_{xy}^0$, while blue and red lines mark the AHC for  positive and negative chirality.
		(g-h) The symmetric, $\sigma_{xy}^{s}$ (violet line), and antisymmetric, $\sigma_{xy}^{a}$ (orange line) parts of the AHC are shown on the background of the AHC in the collinear state (shaded area).
		While the crystal Hall effect ($\sigma_{xy}^{s}$) of SRO displays little variation with the canting plane, the chiral Hall effect ($\sigma_{xy}^{a}$) is extremely sensitive to the interplay of crystal symmetries and canting.
	}
	\label{FIG4}
\end{figure*}

\noindent
{\bf Chiral Hall effect in Mn$_2$Au.} To demonstrate the close relation of the chiral Hall effect to the staggered mixed Berry curvature, we consider an example of Mn$_2$Au. We investigate the AFM phase of this material with the spins on two Mn sublattices (A and B) aligned along the $z$-axis, see Fig.~4(a), and compute its electronic structure and transport properties by referring to ab-initio methods. The crystal structure of Mn$_2$Au possesses global inversion symmetry which prohibits the emergence of the crystal Hall effect, in accordance to the symmetry analysis presented above. The collinear AFM state of this system has $PT$-symmetry which results in degeneracy of the bands for collinear spin configuration (black dashed lines in Fig. 4c).

To simulate the effect of canting, we apply an exchange field of various magnitude along $y$, acting on the set of ab-initio Wannier states. This results in canting of spins on two sublattices in the $zy$-plane, Fig.~4a. The magnitude of the exchange field of $\pm 100$\,meV corresponds to about $\pm 2^{\circ}$ of canting away from the $z$-axis. Finite spin canting and corresponding finite chirality break the $PT$-symmetry, which results in lifting of band degeneracies at each $k$-point in the Brillouin zone, as exemplified for the case of $+ 2^{\circ}$ canting in Fig.~4c. Upon canting, each of the split bands acquires a finite $k$-space Berry curvature $\Omega^n_{zx}(\mathbf{k}) =2\Im \Braket{\partial_{k_z}u_{n\mathbf{k}}|\partial_{k_x}u_{n\mathbf{k}}}$, Fig.~4(c), which is purely anti-symmetric in nature: i.e. upon canting of the opposite sense, while the band structure remains intact, $\Omega^n_{zx}(\mathbf{k})$ retains its magnitude but switches its sign. This means, that in case of Mn$_2$Au, the Hall conductivity, obtained by summing up positive and negative Berry curvature contributions over all bands, Fig.~4c, is manifestly chiral
in that it switches sign with changing the sense of canting. The corresponding computed chiral Hall conductivity, shown as a function of band filling and strength of canting in Fig.~4b, displays a complex structure with pronounced peaks and sizeable magnitude.

To clearly reveal the geometric origin of the chiral Hall effect in Mn$_2$Au along the lines of Berry phase theory  presented above, we calculate the band-resolved  contributions to the staggered mixed Berry curvature $\Omega^n_{\lambda x}(\mathbf{k})=2\Im\Braket{\partial_{k_x}u_n|\partial_{\lambda}u_n}$, where $\lambda$ corresponds to staggered canting by angle $\theta$ of the spins on two sublattices in $zx$-plane. At each $k$-point, the  Berry curvature $\Omega^n_{\lambda x}(\mathbf{k})$, calculated in the collinear AFM state and shown in Fig.~4d, has identical values for the pairs of $PT$-symmetric bands, which is in contrast to the mixed Berry curvature corresponding to the coherent rotation of spins: as result of $PT$ symmetry the mixed Berry curvature and corresponding damping-like spin-orbit torque vanish when summed up over pair of $PT$-symmetric bands~\cite{Hanke_mixed_2017-2,PhysRevLett.118.106402,PhysRevLett.113.157201}. As a result, while the non-staggered damping-like torques are inactive in $PT$-symmetric AFMs such as Mn$_2$Au, the staggered damping-like torques~\cite{PhysRevB.89.174430,PhysRevB.73.214426}, for each state  proportional to  $\Omega^n_{\lambda x}(\mathbf{k})$ but acting in an opposite way on spins in  A and B sublattices, are allowed and can be prominent (see also Discussion section).

By comparing Fig.~4c and d, we observe a very close correlation between the chiral Hall effect and the staggered mixed Berry curvature. We thus numerically solidify the outcome of Eq.~9, which states that large contributions in  $\Omega^n_{\lambda x}$  reflect directly on the magnitude of the chiral Hall conductivity. This correlation is particularly prominent in the vicinity of near  degeneracies among the bands where large contributions to the staggered Berry curvature and chiral Hall conductivity arise. While such degeneracies in Mn$_2$Au often carry an isolated monopole character, such as e.g. at $+$1.7\,eV around X or at $+$1.4\,eV along $\Gamma$M, they also occur along  ``hot" sheets of whole  bands coming close to each other~\cite{PhysRevLett.106.117202}, as is the case for example along AZ, Fig.~4d.  
The finding of the relation between the chiral Hall effect and staggered mixed Berry curvature $-$ and thus staggered damping-like spin-orbit torque $-$ is important as it provides a guiding principle in the material design of both phenomena, and allows to relate the observations of the Hall signal to the physics of spin-orbit torques and vice versa. 

\noindent
{\bf Chiral Hall effect in SrRuO$_3$.}
We now move on to a specific material example which, upon doping, hosts pronounced crystal and chiral Hall effects at the same time.
Namely, we consider a monolayer of SrO-terminated SrRuO$_3$ (SRO) thin films grown on SrTiO$_3$, comprising  two Ru spin moments which are arranged antiferromagnetically in the collinear ground state~\cite{PRL-SRO2020,Xia-PRB2009,Chang-PRL2009,Toyota-MIT2005,Kartik}, with $\hat{\mathbf{s}}_0$ along the $x$-axis in the plane of the film ($xy$-plane), see (Fig. 5 a). In the ground state, the monolayer of SRO exhibits a symmetry breaking associated with rotation and tilts of oxygen octahedra surrounding Ru atoms~\cite{Kartik}.
The band structure of SRO monolayer around the Fermi energy is dominated by Ru-t$_{2g}$ states. The combined effect of octahedral distortion, SOI and AFM ordering on  Ru-t$_{2g}$ states leads to a formation of a 0.96\,eV gap at the Fermi energy and breaking of degeneracies among the bands present in a symmetric phase of this material, see (Fig. 5 b)~\cite{Kartik}. 
The corresponding band splittings are found to be quite prominent around the energies of $-$0.60, $-$0.21 and $+$1.13\,eV, reflecting the strong effect of SOI on the states there, (Fig. 5 b).

Starting from the collinear AFM ground state of the system
we consider a small canting of staggered spins away from the $x$-axis by $\theta= 5^{\circ}$ (chirality ``$+$") and $\theta= -5^{\circ}$ (chirality ``$-$"), both in the $xy$-plane (i.e. keeping the spins in-plane), as well as in the $xz$-plane (as in Fig. 5 a), showing the corresponding rearrangements of the bands for the $xy$ canting plane in Fig. 5 b.
The asymmetric effect of the canting on the electronic band structure is most prominent around the energies of $-$0.21, $-$0.60 and $+$1.13 eV, where the effect of SOI is strongest. 
Here, depending on chirality and  specific Bloch vector, the initial splitting between the ``collinear" Ru-states gets several times larger upon canting.

Next, we assess the intrinsic Berry curvature contribution to the AHE in SRO upon canting and compare it to the AHE in the collinear state (see section Methods for more details).
As was shown recently~\cite{Kartik}, in the collinear (along $x$) state  considered here SRO monolayer exhibits a significant crystal Hall effect over wide regions of energy as a result of combined breaking of  time reversal symmetry 
and translation by half a lattice constant arising as a consequence of octahedral distortion. 
In addition to the crystal Hall conductivity at zero canting, $\sigma_{xy}^0$, shown in (Fig. 5 c,d) with a shaded area, the canting by 5$^\circ$ with positive and negative  chirality induces significant changes to the AHC, irrespective of whether the canting is performed in the $xy$- (Fig. 5 c, top) or $xz$-plane (Fig. 5 d, top). Despite a relatively modest effect on the re-distribution of the bands, the effect of small canting on the AHC is especially drastic in the regions of energy of $[-0.6,-0.5]$ and $[+1.0,+1.2]$\,eV, where the 
magnitude of $\sigma_{xy}^0$  gets significantly enhanced by canting, and its sign depends on chirality. 
We decompose the computed AHC of the canted system into symmetric and antisymmetric components, $\sigma_{xy}^s$ and $\sigma_{xy}^a$, presenting the results in the bottom panels of (Fig. 5 c,d). We clearly observe that for the small canting angle of 5$^\circ$ the crystal Hall conductivity $\sigma_{xy}^s$ follows the energy-dependence of $\sigma_{xy}^0$ quite closely for both tilting planes, which is consistent with the perturbation theory arguments. 

On the other hand, the behavior of the chiral Hall conductivity $\sigma_{xy}^a$ stands in sharp contrast to that of $\sigma_{xy}^s$ and $\sigma_{xy}^0$.
In analogy to Mn$_2$Au, given the smallness of canting, the magnitude of the chiral Hall effect that we observe appears gigantic, and it can be attributed to near band degeneracies, with the cross-talk among them  activated by canting via staggered mixed Berry curvature mechanism. 
While all three types of conductivities originate in the same regions in energy associated with pronounced influence of SOI on the electronic structure, there is no correlation in the sign of $\sigma_{xy}^a$ and $\sigma_{xy}^s$, and the peaks in $\sigma_{xy}^a$ are often not correlated with the sharp features of crystal Hall effect, which is particularly visible for the case of $xz$-canting. 
This is consistent with the picture that the states which give rise to the chiral Hall effect and which are  sensitive to the canting-driven symmetry breaking are not necessarily associated with the crystal $-$ i.e. "conventional" anomalous $-$ Hall effect.
The comparison of $\sigma_{xy}^a$ for two different canting planes,~(Fig. 5 c,d), reveals extreme sensitivity of the chiral Hall effect to the crystal symmetry of the lattice. 
In this sense, tracking the chiral Hall effect with respect to two independent planes of canting provides us with a detailed information on the underlying crystal symmetry without the need of changing the ground state direction of staggered magnetization.
\begin{figure}[t!]
	\begin{center}
		\rotatebox{0}{\includegraphics[width=0.50\textwidth]{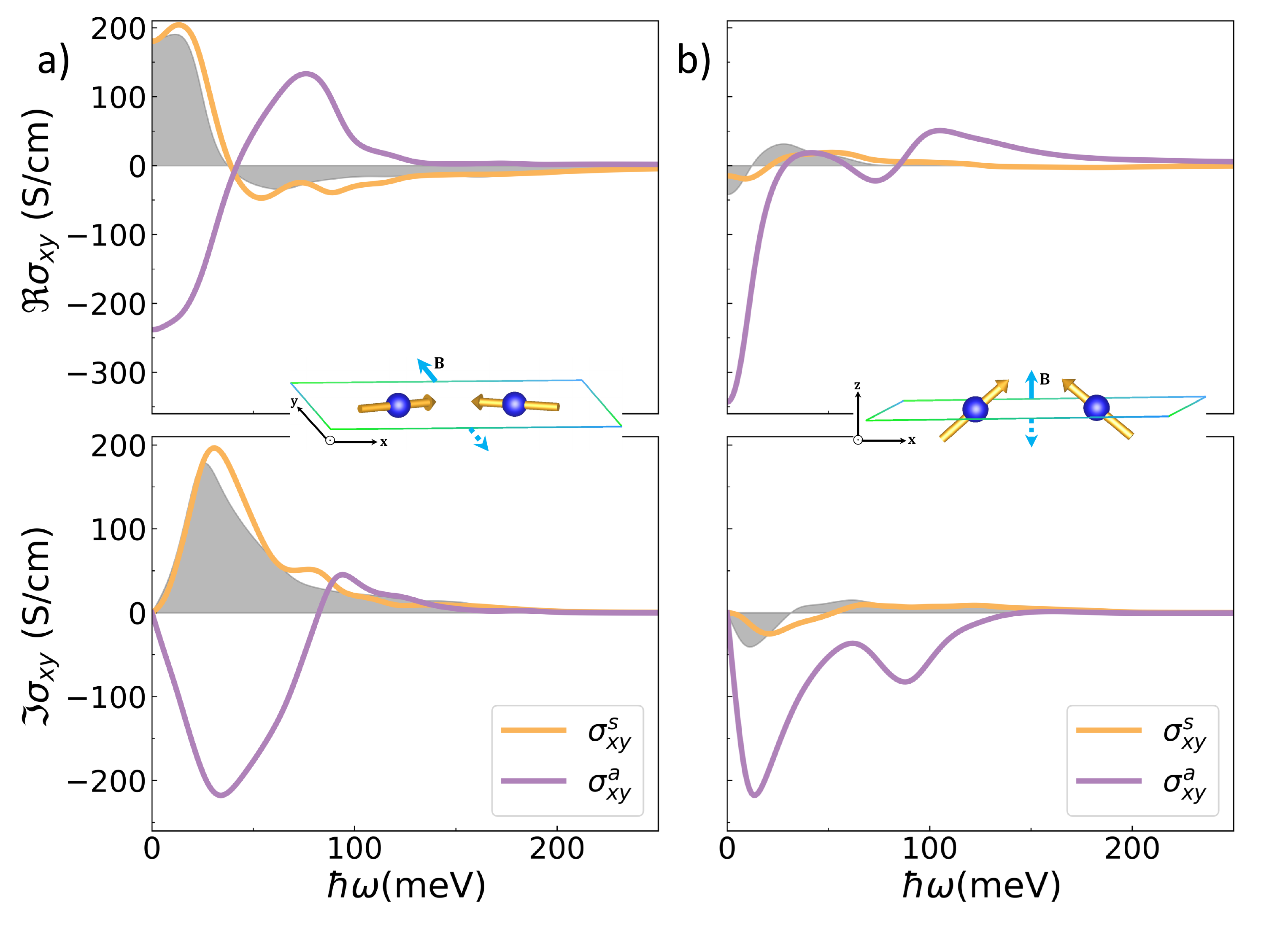}}
	\end{center}
	\caption{{\bf Chiral magneto-optical effect in monolayer of antiferromagnetic SrRuO$_3$}. a) Real (top) and imaginary (bottom) part of the  magneto-optical conductivity in the collinear state (grey shaded area) as well as its symmetric and antisymmetric parts for 5$^\circ$ spin canting in the $xy$-plane evaluated for the position of the Fermi energy at $E_F=1.05$\,eV. (b) Same as in (a) but for the $xz$-plane of canting evaluated at $E_F=1.01$\,eV. The sketches 
		depict the canting of the spins in antiferromagnetic SRO monolayer upon an application of an external magnetic field along the $\pm y$-axis (c), and $\pm z$-axis (d).}
	\label{FIG5}
\end{figure}

\noindent
{\bf Chiral Magneto-Optical Effect.} Finally, we show that the chiral contributions arise not only in the context of the AHE, but also in the realm of magneto-optical (MO) effects. In order to do this, we numerically evaluate the real and imaginary parts of the magneto-optical conductivity (see Methods section for details) of monolayer SRO, starting from the collinear AFM configuration. We further define the symmetric and antisymmetric parts of the magneto-optical conductivity, $\sigma_{xy}^s(\omega)$ and $\sigma_{xy}^a(\omega)$, by referring to the frequency-dependent version of Eq.~\ref{Eq1}, upon canting by $5^{\circ}$ of opposite chirality in the $xy$- and $xz$-plane, in analogy to the previous section.

The results of our assessment are presented in Fig. 6, where we have chosen the Fermi energy to be positioned at the peak of the chiral Hall effect for the corresponding rotation plane as shown in (Fig. 6 c,d): at $E_F=1.05$\,eV for $xy$-, and at $E_F=1.01$\,eV for $xz$-plane of canting. Our analysis shows that, in analogy to their~d.c.~versions, the crystal magneto-optical conductivity follows quite closely the frequency distribution of the MO  conductivity computed without canting, both in its real and imaginary parts. On the other hand, while the magnitude of chiral MO conductivity remains large over a wide region of frequencies, its structure is often not  correlated
with the corresponding behavior of the crystal part of the conductivity in $\omega$: for example in case of $xz$-canting the chiral MO conductivity is very prominent on the background of almost vanishing crystal MO conductivity in the entire range of energies. This marks the two effects as distinct magneto-optical phenomena. The chiral MO effect thus presents a unique tool to track down optically-mediated electronic transitions 
which are responsive to the effect of canting. Tracing down the chiral contributions to the MO conductivity makes it possible to gain a valuable insight into the interplay of electronic structure with crystal symmetry and magnetic order.

\vspace{0.5cm}
\noindent
%{\large{\bf Discussion}}\\
\section*{Discussion}
In this work, we promote the chiral Hall effect as a new tool to access the properties of ferromagnetic and antiferromagnetic materials. We uncovered that the chiral Hall effect has a qualitatively different Berry phase origin as compared to the conventional AHE. Based on this, we are able to understand how a gigantic chiral Hall effect can be achieved in compensated AFMs even upon a very small canting accompanied by an almost vanishing ferromagnetic component of the magnetization. Addition of the chiral Hall effect to the crystal Hall effect thus allows for drawing a unified map of Hall effects taking place in canted magnets.

As we have seen on the example of SrRuO$_3$, the chiral Hall effect is 
sensitive to the details of crystal structure, depending on the plane of canting. In a realistic situation, given a robust ground state direction of the staggered magnetization $\hat{\mathbf{s}}_0$ in an AFM, which is accompanied by a vanishing or non-vanishing crystal Hall effect, the plane of canting can be straightforwardly  controlled by a direction of an externally applied magnetic field $\mathbf{B}$, and the chiral Hall effect can be estimated as a difference in measured Hall effect between opposite directions of $\mathbf{B}\rightarrow -\mathbf{B}$, see sketches in (Fig. 5). Sweeping the direction of the field in the plane orthogonal to $\hat{\mathbf{s}}_0$ would allow to reconstruct the angular dependence of the chiral Hall conductivity and determine its nodal points (i.e.~the directions for which it turns to zero), from which the information about the details of the crystal symmetry can be deduced.
On the other hand, the response of the measured signal to the strength of the magnetic field can be used to estimate the magnitude of the Berry curvature response as given by the geometrical theory, Eq.~(\ref{mixed}). The corresponding experimental assessment of the evolution of chiral magneto-optical conductivity, in combination with the magneto-optical spectra without the field, can be used to reconstruct the exact details of electronic structure of a given material, especially the energetic position of states  sensitive to canting that hosts large staggered mixed Berry curvature. 

Although the role of the chiral Hall effect in ferromagnets is more difficult to access as it is difficult to realize the states of opposite chirality in analogy to AFMs (especially in systems with collinear ground state), the chiral Hall effect, as the dominant contribution to the variation of the AHE upon canting, can contribute strongly to the evolution of the AHE with temperature via the effect of fluctuations. This is easy to understand by realizing that even in collinear ferromagnets with DMI the temperature fluctuations will promote one type of chirality over the other~\cite{Menzel}, which will prohibit the opposite contributions to the AHE from the states of opposite chirality from suppressing each other.
The variation of the AHE with temperature $T$ corresponding to the chiral Hall effect is expected to behave qualitatively differently with respect to the temperature-induced magnetization change $\Delta M(T)$, which at low $T$ is proportional to $\theta^2$ with $\theta(T)$ being an effective fluctuations-driven deviation of the local spins from the equilibrium magnetization direction. Indeed, while the conventional theory of the AHE assumes that the variation of the anomalous Hall resistivity with $T$ is proportional to $\Delta M(T)$ and thus to $\theta^2(T)$, the chiral Hall effect imposes a different, linear in $\theta(T)$ behavior. The fingerprints of the chiral Hall effect can be thus uncovered from the scaling analysis of the temperature-dependent Hall measurements in FM  materials.

A promising approach to induce  canting between collinear spins in a ferromagnetic ground state, and thereby ignite the chiral Hall effect, lies in referring to current-induced staggered spin-orbit torques (SOTs) ~\cite{RMP-SOT}, which have shown above to be closely linked to the microscopics of the chiral Hall conductivity.
Given that an electric field applied to a ferromagnet exerts local torques on the spins $\mathbf{T}_{\rm A}$ and $\mathbf{T}_{\rm B}$, a crucial distinction can be drawn. While the non-staggered conventional SOT,  $\mathbf{T}_+=\mathbf{T}_{\rm A}+\mathbf{T}_{\rm B}$ leads to a coherent magnetization rotation~\cite{RMP-SOT,Frank-SOT,Miron,Wadley}, the staggered component of the SOT~\cite{Jakub} defined as $\mathbf{T}_-=\mathbf{T}_{\rm A}-\mathbf{T}_{\rm B}$ additionally will attempt to induce a finite canting in the system.
In analogy to $\mathbf{T}_+$~\cite{Frank-SOT}, components of staggered SOT even and odd with respect to $\mathbf{n}_+$, $\mathbf{T}_-^{\rm even}$ and $\mathbf{T}_-^{\rm odd}$, can be distinguished.  
In systems with inversion symmetry staggered polar tensors of even rank and staggered axial tensors of odd rank are forbidden by symmetry which means that  $\mathbf{T}_-^{\rm even}$ and $\mathbf{T}_-^{\rm odd}$ are even in the Rashba strength. However, polar tensors of odd rank and axial tensors of even rank are forbidden by symmetry, and consequently both components of $\mathbf{T}_+(\mathbf{E})$ are odd in $\alpha_\text{$\rm R$}$. Therefore, in contrast to non-staggered SOT, the staggered torques in ferromagnets do not necessarily require broken inversion symmetry. Staggered SOTs can be also used to induce canting in collinear AFMs~\cite{RMP-SOT}, in which case one has to distinguish components which are even and odd with respect to $\mathbf{n}_-$.  As $\mathbf{T}_-^{\rm odd}(\mathbf{E})$ is a polar tensor and $\mathbf{T}_-^{\rm even}(\mathbf{E})$  is a staggered axial tensor, it can be shown that
$\mathbf{T}_-^{\rm odd}(\mathbf{E})$ and $\mathbf{T}_-^{\rm even}(\mathbf{E})$ are respectively odd and even in the Rashba strength. 

Generally, the interplay of the chiral Hall effect with current-driven phenomena presents an exciting avenue to explore. 
By referring to the mechanism of staggered torques, the chiral Hall effect can manifest as a non-linear contribution to the Hall effect, in analogy to the non-linear magnetoresistance effect used to detect the Néel vector reversal in collinear AFMs~\cite{Godinho}.
Besides the relation of the chiral Hall effects to various types of spin-torques born in the system when a current passes through it, the new flavor of the Hall effect should be also intertwined with the phenomenon of current-induced DMI, where the sense and magnitude of canting among spins can be altered upon passing a current through the sample~\cite{Karnad,Hayashi}.
Moreover, the correlation of the chiral Hall effect with the modifications in the electronic structure brought by an external electric field~e.g.~in multiferroics materials must be also profound.

In our work, we have defined the crystal and chiral Hall effects with respect to the staggered ($\mathbf{n}_-$) and ferromagnetic ($\mathbf{n}_+$) components for the system consisting of two spins, which ultimately allowed for representation in terms of the vector chirality. The generalization of this approach to multi-spin systems, for example Mn$_3$X type of systems~\cite{Nayak2016,Jakub-2017}, B20~\cite{Sergii}, FeMn-type~\cite{Wanxiang-2} or Heusler compounds~\cite{Heusler} presents an exciting challenge. In the latter cases, the symmetry properties of the anomalous Hall effect can be scrutinized with respect to generalized AFM order parameters. In analogy to this work, different flavors of spin and structural chirality can be singled out, and their role in mediating various contributions to the AHE can be identified. Ultimately, the classification obtained from such an analysis, of which our study presents a starting toy case, could be possibly reinterpreted in terms of quantitave and qualitative multipole theory~\cite{multipoles,Oppeneer}, and relation to various types of current-induced phenomena, such as spin torques, could be established. We believe this general direction of research to be of fundamental and practical importance to our understanding of chiral magnetism and our ability to detect and control various chiral magnetic phases and their dynamics.   

\vspace{0.5cm}

\noindent
%{\large{\bf Methods}}
\section*{Methods}
\noindent
{\bf Tight-binding model calculation.}
From tight-binding Hamiltonian the Berry curvature was calculated according to the standard expression
$
\Omega_{n}(\mathbf{k}) = -\hslash^{2} \sum_{n \neq m} [\operatorname{2 Im} \langle u_{n\mathbf{k}}| \hat{ v}_{x}|u_{m\mathbf{k}}\rangle \langle u_{n\mathbf{k}}|\hat{v}_{y}|u_{m\mathbf{k}}\rangle]/(\varepsilon_{n\mathbf{k}}-\varepsilon_{m\mathbf{k}})^{2},
$
where $\Omega_{n}(\mathbf{k})$ is the Berry curvature of band $n$, $\hslash\hat{v}_{i} ={\partial \hat{H}(\mathbf{k})}/{\partial k_{i}} $ is the $i$'th velocity operator, $u_{n\mathbf{k}}$ and $\varepsilon_{n\mathbf{k}}$ are the eigenstates and eigenvalues of the Hamiltonian $\hat{H}(\mathbf{k})$, respectively. From 2 to 4 million $k$-points in the full BZ were used to arrive at well-converged values of the  anomalous Hall conductivity (AHC) determined as 
$\sigma_{x y}= -\hbar e^{2} \int_{BZ} \frac{d\bf k}{(2 \pi)^{2}}  \Omega(\mathbf{k})$,
where $\Omega(\mathbf{k})$ is the sum (for each k) of Berry curvatures over the occupied bands. 
All calculations were done at $T=10$\,K. 

\noindent
{\bf First-principles calculation of Mn$_2$Au.} Electronic structure of Mn$_2$Au was calculated by using a density functional theory (DFT) code \texttt{FLEUR}~\cite{fleur}, which implements the full-potential linearized augmented plane wave (FLAPW) method. Exchange and correlation effects were included within the generalized gradient approximation (GGA) by using Perdew-Burke-Ernzerhof (PBE) functional~\cite{pbe} exchange-correlation functional. Lattice constants of a tetragonal cubic unit cell were set $a=6.29a_0$ and $c=16.14 a_0$, where $a_0$ is the Bohr radius. The muffin-tin radii of Mn and Au were chosen to be $2.53 a_0$ for both atoms. Plane wave-cutoff was set $3.9 a_0^{-1}$, and the BZ was sampled on $12\times 12\times 12$ Monkhorst-Pack ${k}$-mesh~\cite{Monkhorst-Pack1976}. 

To calculate the Berry curvature, we obtained a tight-binding model of Mn$_2$Au by projecting Bloch functions onto 18 initial guess Wannier functions (WFs) -- $s$, $p$, $d$ orbitals with spin up and down -- for both Mn and Au atoms and obtained maximally-localized WFs (MLWFs)~\cite{Frank-WFs,Pizzi_2020}. To induce spin canting, we additionally included an exchange field along $y$ by $H_\mathrm{XC}=(J_\mathrm{XC}/\hbar)S_y$. The Berry curvature shown in Fig.~\ref{FIG4}(c) was calculated by
\begin{eqnarray}
\Omega_{zx}^n (\mathbf{k})
&=&
-2 \hbar^2 
\frac{
\mathrm{Im}
\left[
\bra{u_{n\mathbf{k}}} \hat{v}_z \ket{u_{m\mathbf{k}}}
\bra{u_{n\mathbf{k}}} \hat{v}_x \ket{u_{m\mathbf{k}}}
\right]
}{ (\varepsilon_{n\mathbf{k}} - \varepsilon_{m\mathbf{k}})^2+\eta^2}
\end{eqnarray}
where we set $\eta=25\ \mathrm{meV}$. The AHC shown in Fig.~\ref{FIG4}(b) was obtained by integrating the Berry curvature over $240\times 240 \times 240$ $k$-points for occupied states. The staggered mixed Berry curvature shown in Fig.~\ref{FIG4}(d) was evaluated by
\begin{eqnarray}
\Omega_{\lambda x}^n (\mathbf{k})
&=&
-2 \hbar
\frac{
\mathrm{Im}
[
\bra{u_{n\mathbf{k}}} \partial_\lambda \hat{H} \ket{u_{m\mathbf{k}}}
\bra{u_{n\mathbf{k}}} \hat{v}_x \ket{u_{m\mathbf{k}}}
]
}{ (\varepsilon_{n\mathbf{k}} - \varepsilon_{m\mathbf{k}})^2+\eta^2},
\end{eqnarray}
where $\lambda$ is defined as a canted angle of the magnetic moments on Mn-A and Mn-B in $zx$ plane. Note that it is related by a staggered torque operator
\begin{eqnarray}
\partial_\lambda \hat{H}
&=&
\frac{\partial \hat{H}}{\partial \theta^\mathrm{A}}
-
\frac{\partial \hat{H}}{\partial \theta^\mathrm{B}}
\nonumber
\\
&=&
\frac{1}{i\hbar}
\left[
\hat{S}_y^\mathrm{A} - \hat{S}_y^\mathrm{B}, \hat{H}
\right]
\nonumber
\\
&=&
\hat{T}_y^\mathrm{A}
-\hat{T}_y^\mathrm{B}
\end{eqnarray}
where $\hat{S}_y^A$ and $\hat{S}_y^B$ are spin operators on Mn-A and Mn-B atoms, respectively.

\noindent
{\bf First-principles calculation of SrRuO$_3$.} DFT calculations were carried out with the FLAPW method as implemented in the \texttt{FLEUR} code~\cite{fleur}.
Using relaxed atomic positions of the SRO monolayer, the electronic structure calculations at different spin canting were carried out with the film version of the FLEUR code~\cite{fleur}.
For self-consistent calculations with the LAPW basis set a plane-wave cutoff of $k_{max}= 4.2a_0^{-1}$ and the total of 24$\times$24 $k$-points in the BZ were used for the convergence of the charge density.
The muffin-tin radii for Sr, Ru, O were set to \SI{2.80}{\au}, \SI{2.32}{\au}, and \SI{1.31}{\au}, respectively.
We used the PBE~\cite{pbe} exchange-correlation functional within the GGA. The electron-electron correlation effects beyond GGA at the magnetic Ru ions were taken into account by referring to the GGA+$U$ method as implemented in the SPEX code~\cite{spex1}, resulting in Coulomb interaction strength of $U=2.52$\,eV and an intra-atomic exchange interaction strength of $J=0.44$\,eV.

To compute the Berry curvature, we first constructed a tight-binding Hamiltonian in terms of maximally-localized Wannier functions 
projected from the GGA+$U$+SOC\,[100] states using atomic-orbital-like Ru-t$_{2g}$ and Ru-e$_{g}$ states as initial guess~\cite{Frank-WFs,Pizzi_2020}. From this Hamiltonian the Berry curvature is calculated on a $50\times 50$ $k$-mesh employing an adaptive $5\times 5$ refinement scheme~\cite{Wang-Souza} at points where the value of the Berry curvature exceeded 50\,a.u. These numerical parameters provided well-converged values of the  anomalous Hall conductivity.
The magneto-optical conductivity was calculated using the Kubo expression
\begin{equation}
\begin{aligned}
\sigma_{x y}(\omega)=& \hbar e^{2} \int \frac{d \mathbf{k}}{(2 \pi)^{2}} \sum_{n \neq m}\left(f_{n \mathbf{k}}-f_{m \mathbf{k}}\right) \\
& \times \frac{\operatorname{Im}\left[\left\langle u_{n \mathbf{k}}\left|\hat{v}_{x}\right| u_{m \mathbf{k}}\right\rangle\left\langle u_{m \mathbf{k}}\left|\hat{v}_{y}\right| u_{n \mathbf{k}}\right\rangle\right]}{\left(\varepsilon_{n \mathbf{k}}-\varepsilon_{m \mathbf{k}}\right)^{2}-(\hbar \omega+i \eta)^{2}},
\end{aligned}
\end{equation}
where    $\hbar\omega$ is the frequency of the applied electric field, and $\eta$ a material dependent broadening parameter. For calculations presented in (Fig. 5) we used  $\eta=10$\,meV.

\vspace{0.5cm}

\noindent
\section*{Acknowledgements}
We thank Libor \v{S}mejkal for extensive discussions on the subject. We  acknowledge  funding  under SPP 2137 ``Skyrmionics" of the DFG.
We gratefully acknowledge financial support from the European Research Council (ERC) under the European Union's Horizon 2020 research and innovation program (Grant No. 856538, project "3D MAGiC”).
We  also gratefully acknowledge the J\"ulich Supercomputing Centre and RWTH Aachen University for providing computational resources under project Nos. jiff40 and jpgi11. 
The work was also supported by the Deutsche Forschungsgemeinschaft (DFG, German Research Foundation) $-$ TRR 173 $-$ 268565370 (project A11), TRR 288 – 422213477 (project B06), and project MO 1731/10-1 of the DFG. We also acknowledge funding under HGF-RSF Joint Research Group ``TOPOMANN". 
\section*{Competing Interests}
The authors declare no competing interests.
\section*{Data Availability Statement}
The data that support the findings of this study are available from the corresponding author upon reasonable request.
%----------------------------------------------------
% literature
%----------------------------------------------------

\hbadness=99999
\bibliographystyle{naturemag}%{Science.bst}
\bibliography{literature}
% %----------------------------------------------------
%				Literature end
% %----------------------------------------------------
\end{document}

\setcounter{secnumdepth}{2} 

%----------------------------------------------------
% Document title
%----------------------------------------------------

\title{Supplemental Material: The chiral Hall effect in canted ferromagnets and antiferromagnets}

\author{J. Kipp}
    \affiliation{\pgi}
    \affiliation{\aachen}
\author{K. Samanta}
    \affiliation{\pgi}

\author{F. R. Lux}
    \affiliation{\pgi}
    \affiliation{\aachen}

\author{M. Merte}

    \affiliation{\pgi}
    \affiliation{\aachen}
    \affiliation{\mainz}
    
\author{D. Go}
   \affiliation{\mainz}
    \affiliation{\pgi}

\author{J.-P. Hanke}

    \affiliation{\pgi}

\author{M. Redies}

    \affiliation{\pgi}
    \affiliation{\aachen}
\author{F. Freimuth}

    \affiliation{\pgi}
    
\author{S. Bl\"ugel}

    \affiliation{\pgi}
    
\author{M. Le\v{z}ai\'c}

    \affiliation{\pgi}
    
\author{Y. Mokrousov}

    \affiliation{\pgi}
    \affiliation{\mainz}

%%%%%%%%%%%%%%%%% END OF PREAMBLE %%%%%%%%%%%%%%%%

\maketitle

%----------------------------------------------------
% Document body
%----------------------------------------------------
\date{\today}

%----------------------------------------------------
% literature
%----------------------------------------------------

\hbadness=99999
\bibliographystyle{naturemag}%{Science.bst}
\bibliography{literature}
% %----------------------------------------------------
%				Literature end
% %----------------------------------------------------

\appendix
\renewcommand\thefigure{\arabic{figure}}%\thesection
%%%REDEFINE SECTION COMMMAND TO HAVE NAME == SUPPLEMENTARY NOTE 1,2,3
\renewcommand{\thesection}{\arabic{section}}
\titleformat{\section}
{\normalfont\bfseries\filcenter}{Supplementary Note~\thesection:}{0.2em}{}
\titleformat{\subsection}
{\normalfont\bfseries\filcenter}{\Alph{subsection}:}{0.2em}{}
%\titleformat{<command>}[<shape>]{<format>}{<label>}{<sep>}
%%%%%%%%%%%%%%%%%%%%
\renewcommand{\figurename}{Supplementary Fig.}

\section{ Symmetry analysis of the crystal and chiral Hall effect for the hexagonal magnetic Rashba model}%Supplemental Note 1:

The Rashba model on the hexagonal lattice which is studied in the manuscript belongs to the crystallographic point group $C_{6v}$ in the non-magnetic case.
When the system is magnetic, it is characterized by the respective magnetization vectors $\vec{s}_{i}$ at each of the sites $i=A,B$.
The direction cosine expansion in the manuscript is expressed in terms of $\vec{n}_\pm = \vec{s}_\mathrm{A} \pm \vec{s}_\mathrm{B} $, i.e,
\begin{align}
\sigma_{xy}^\mathrm{odd} &= \sum_{k,l=0}^\infty (c^\mathrm{odd}_{xy})^i :  ( \vec{n}_-^{\otimes 2k+1}   \otimes  \vec{n}_+^{\otimes 2l} )_i
\label{eq:sigma_expansion_A}
\\
\sigma_{xy}^\mathrm{even} &= \sum_{k,l=0}^\infty (c^\mathrm{even}_{xy})^i :  ( \vec{n}_-^{\otimes 2k}   \otimes  \vec{n}_+^{\otimes 2l+1} )_i .
\label{eq:sigma_expansion_B}
\end{align}
Accodringly, the lowest orders are thus given by the general expansions
\begin{align}
\sigma_{xy}^\mathrm{odd} &= 
  (c^\mathrm{odd}_{xy})^i (n_-)_i + (c^\mathrm{odd}_{xy})^{ikl} (n_-)_i (n_+)_k (n_+)_l + \mathcal{O}( n^5) 
  \label{eq:odd_trafo}
  \\
  \sigma_{xy}^\mathrm{even} &= 
  (c^\mathrm{even}_{xy})^i (n_+)_i + (c^\mathrm{even}_{xy})^{ikl} (n_+)_i (n_-)_k (n_-)_l + \mathcal{O}( n^5) .
  \label{eq:even_trafo}
\end{align}
If a physical tensor $T$ of rank $n=k+l$ connects $k$ polar and $l$ axial vectors, its transformation law under a symmetry transformation $S$ is usually given by
\begin{equation}
    T^{p_1 \cdots p_k a_1 \cdots a_l} ~~\rightarrow~~
T^{p_1 \cdots p_k a_1 \cdots a_l}  |S|^l 
S^{\tilde{p}_1}_{p_1}\cdots S^{\tilde{p}_k}_{p_k}
S^{\tilde{a}_1}_{a_1}\cdots S^{\tilde{a}_l}_{a_l},
\end{equation}
where $|S| = \det S$.
Whether $\hatn_\pm$ transforms as a polar or axial vector depends however on the position of the magnetic atoms and which symmetry transformations interchange the lattice sites.
For example, while $ P \hatn_{A,B} = P \hatn_{A,B}$, one has  $P \hatn_{-} = -  \hatn_{-}$ if $P$ interchanges $A$ and $B$, which would be the case for magnetic atoms on the hexagonal lattice.
In other words, $\hatn_{-}$ behaves as a polar vector under $P$ while $\hatn_+$ behaves as an axial vector.
Taking the possible interchange of magnetic lattice sites into account for an arbitrary symmetry transformation $S$, the transformation law is generalized to
\begin{equation}
    T^{i_1 \cdots i_n } ~~\rightarrow~~
T^{i_1 \cdots i_n}   \mathrm{swap}(S)^\eta |S|^\epsilon
S^{\tilde{i}_1}_{i_1}\cdots S^{\tilde{i}_k}_{i_n},
\end{equation}
where $\mathrm{swap}(S) = -1$, if $S$ interchanges the lattice sites $A$ and $B$ and $\mathrm{swap}(S) = 1$ otherwise.
As before, a tensor with $\epsilon = 0$ is called polar and with $\epsilon = 1$ it is axial. 
If $\eta = 1$ instead of $\eta=0$, the tensor is called \emph{staggered}, since its build to compensate the coupling to the staggered fields $n_{-}$ and their unusual behavior under symmetry transformations.
In this terminology, the coefficients of $\sigma_{xy}^\mathrm{odd}$ qualify as staggered axial tensors, while the coefficients of $\sigma_{xy}^\mathrm{even}$ belong to an ordinary axial tensor. 
This fixes their transformation behavior and one finds
\begin{align}
\frac{\sigma_{xy}^\mathrm{odd} - \sigma_{yx}^\mathrm{odd}}{2} & \in \mathrm{span} \lbrace
n_{-}^y n_{+}^y n_{+}^x + n_{-}^y n_{+}^x n_{+}^y
+ n_{-}^x n_{+}^y n_{+}^y - n_{-}^x n_{+}^x n_{+}^x
\rbrace
~ + \mathcal{O}( n^5) \\
 \frac{\sigma_{xy}^\mathrm{even} - \sigma_{yx}^\mathrm{even}}{2} &= 
 \in \mathrm{span} \lbrace
n_{+}^z,
n_{-}^z n_{-}^x n_{+}^x + n_{-}^z n_{-}^y n_{+}^y, ~
n_{-}^x n_{-}^z n_{+}^x + n_{-}^y n_{-}^z n_{+}^y, ~
n_{-}^x n_{-}^x n_{+}^z + n_{-}^y n_{-}^y n_{+}^z
\rbrace~ +  \mathcal{O}( n^5) ,
\end{align}
for the point group $C_{6v}$ (the symmetry of the Rashba model on the hexagonal lattice as presented in the manuscript) and where the $x$-direction is defined as the axis which contains both atoms of the unit cell.
Since $\vec{n}_+ \cdot \vec{n}_- = 0$, one has
$
\boldsymbol{\chi} \times \vec{n}_\pm = \mp \| \vec{n}_\pm \|^2 \vec{n}_\mp /2$, which can be used to introduce the chirality to these results.
One finds 
\begin{align}
n_{-}^y n_{+}^y n_{+}^x + n_{-}^y n_{+}^x n_{+}^y
+ n_{-}^x n_{+}^y n_{+}^y - n_{-}^x n_{+}^x n_{+}^x
= -\frac{2}{\| \mathbf{n}_{+} \|^2} \begin{pmatrix}
- 2n_{+}^x n_{+}^y n_{+}^z\\
( (n_{+}^y)^2-(n_{+}^x)^2) n_{+}^z \\
3 (n_{+}^x)^2 n_{+}^y - (n_{+}^y)^3
\end{pmatrix} \cdot \boldsymbol{\chi}.
\end{align}
This result can be recast into the form 
\begin{equation}
    \frac{\sigma_{xy}^\mathrm{odd} - \sigma_{yx}^\mathrm{odd}}{2}
    \propto \vec{n}_{+}^T   \Xi~ \boldsymbol{\chi}  + \mathcal{O}(\chi^3),
\end{equation}
with the angular-dependent coupling matrix
\begin{equation}
    \Xi
     = -2
    \begin{pmatrix}
    - 2 \hat{n}_{+}^y \hat{n}_{+}^z & 0 & 0 \\
    0 & 0 & ( 3 (\hat{n}_{+}^x)^2 - (\hat{n}_{+}^y)^2 ) \\
    0 & ( (\hat{n}_{+}^y)^2-(\hat{n}_{+}^x)^2) & 0 
    \end{pmatrix},
\end{equation}
and where $\hatn_{+} = \vec{n}_{+} / \| \vec{n}_+ \|$.
In analogy to the topological Hall effect, one might expect the emergence of an isotropic coupling of the form $\vec{n}_+ \cdot \boldsymbol{\chi}$, known as the \emph{scalar spin chirality}.
For the magnetic Rashba model however, this quantity is always zero, i.e., $\vec{n}_+ \cdot \boldsymbol{\chi}=0$.
A finite effect can still arise through anisotropic deviations from this isotropic coupling which are ultimately induced by the spin-orbit interaction. 

\begin{figure}[t]
\centering
\begin{subfigure}{6cm}
  \centering
  \includegraphics[width=\linewidth]{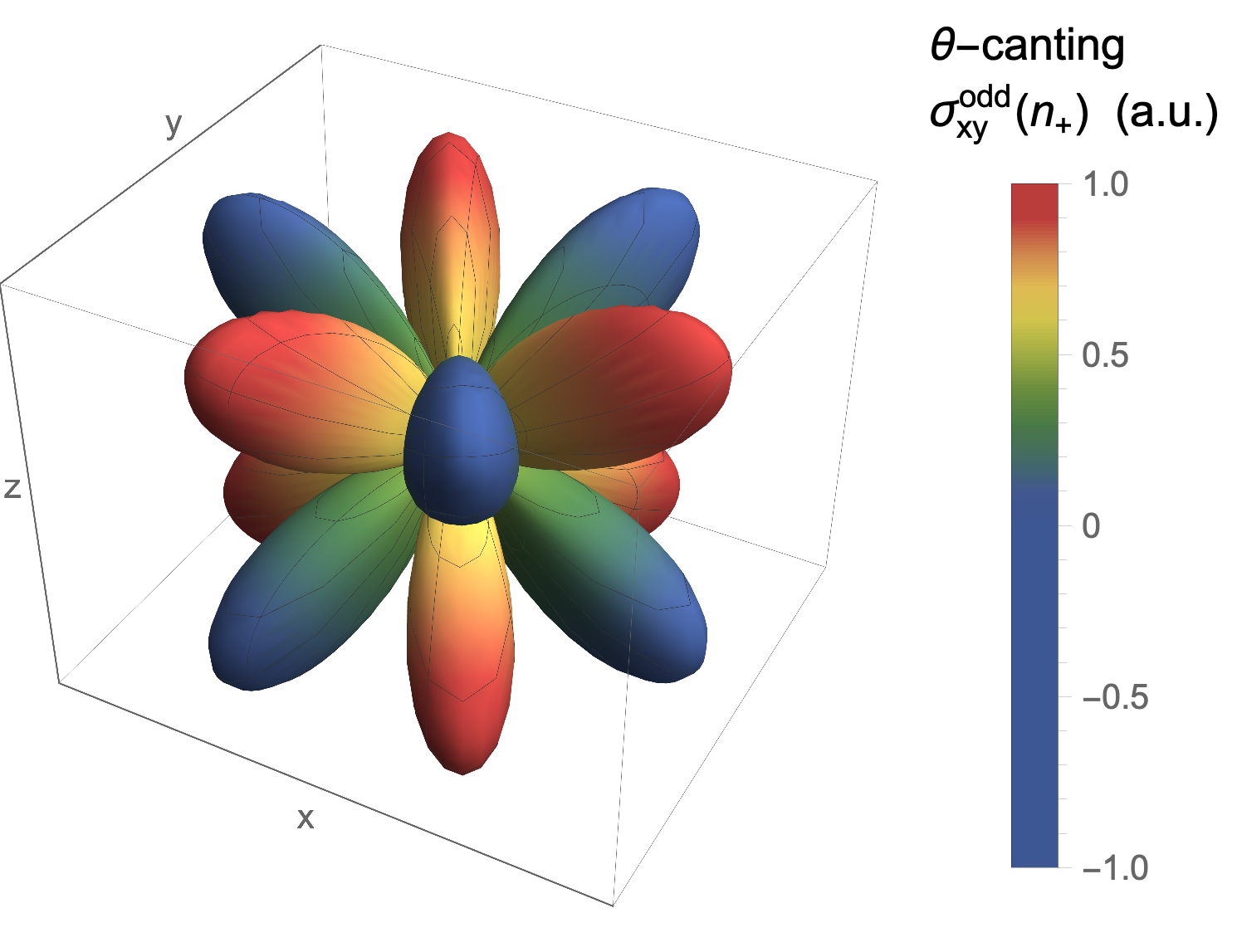}
  \caption{\small Canting with respect to the $\theta$-angle}
  \label{fig:sub1}
\end{subfigure}%
\begin{subfigure}{6cm}
  \centering
  \includegraphics[width=\linewidth]{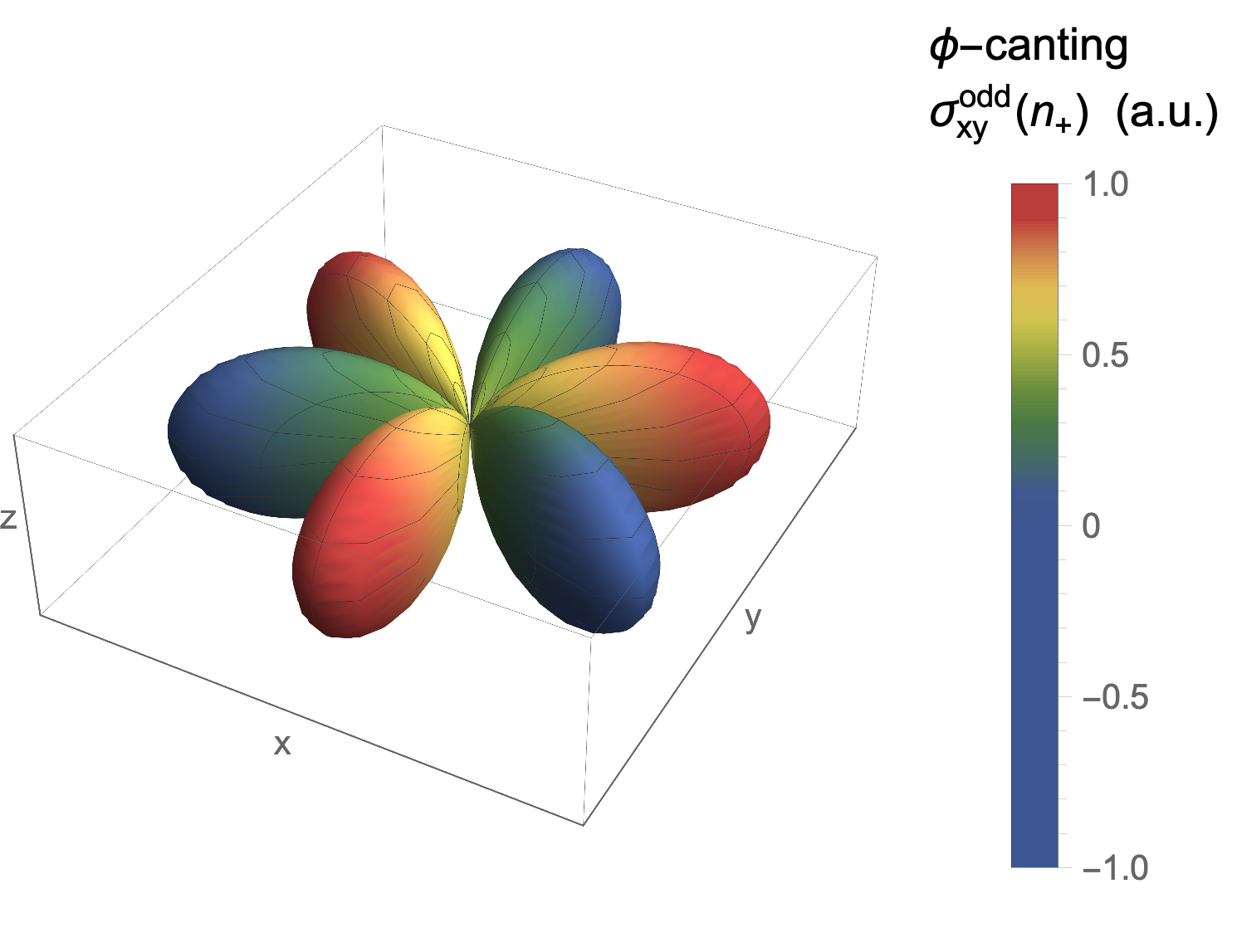}
  \caption{\small Canting with respect to the $\phi$-angle}
  \label{fig:sub2}
\end{subfigure}
\caption{\small Symmetry analysis of the chiral Hall effect in ferromagnets. The two figures show the parametric surface of the antisymmetrized contributions to $\sigma_{xy}^\mathrm{odd}$, parametrized by the direction of the ferromagnetic order parameter $\vec{n}_{+}$.
The antiferromagnetic order parameter is induced by canting the ferromagnetic state either with respect to the polar angle $\theta$, or the azimuthal angle $\phi$.
The $x$-axis is defined to contain the two atoms of the unit cell.
\hspace*{\fill}
}
\label{fig:symmetry_visualization}
\end{figure}

The symmetry of the odd tensor can be visualized by applying the $\theta$-canting procedure as defined in Fig. 2a) of the manuscript.
For $\vec{n}_{-} = \mathrm{const}$, Fig.~\ref{fig:symmetry_visualization}a) explores the angular dependence of the tensor, by varying the spherical angles of $\vec{n}_{+}$ and plotting the magnitude of $(\sigma_{xy}^\mathrm{odd}(\theta)-\sigma_{xy}^\mathrm{odd}(-\theta) )/2$ as a parametric surface.
The result reveals that the effect vanishes when $\vec{n}_{+}$ is contained in the $yz$ mirror plane, i.e., when $n_+^x = 0$.
Likewise, the $\theta$-canting is ineffective when $n^z_+ = 0$.
A dual situation occurs if the canting is not induced via the polar angle $\theta$, but via the azimuthal angle $\phi$.
The parametric surface which corresponds to $(\sigma_{xy}^\mathrm{odd}(\phi)-\sigma_{xy}^\mathrm{odd}(-\phi) )/2$ is shown in Fig.~\ref{fig:symmetry_visualization}b).
In this case, the effect is largest when $n^z_+ = 0$ and it vanishes when $\vec{n}_+$ is contained in the $xz$ mirror plane.

\subsection{Influence of mirror planes}

Since the mirror planes seem to define an important characteristic of the chiral Hall effect on the hexagonal lattice, it is worthwhile to study in detail how they enter the symmetry analysis.
In a first step, one can define the auxiliary variables
\begin{equation}
    m_i = \left\lbrace 
    \begin{array}{cc}
        -1,&  \text{if}~ i = y \\
         1,& \text{otherwise}.
    \end{array}\right. ,
\end{equation}
which are used to specify the $xz$ mirror operator, given by $(M)^i_j = \delta^i_j  m_i$. 
Therefore $| M | = -1$. 
The transformation law  for the coefficients in Eq.~(\ref{eq:odd_trafo}) and Eq.~(\ref{eq:even_trafo}) now reads
\begin{align}
    c_{xy}^{i_1 \cdots i_n } \overset{!}{=}
c_{xy}^{ i_1 \cdots i_n}   \mathrm{swap}(M)^\eta |M|^\epsilon m_x m_y
 \prod_i m_i.
\end{align}
This means that $c_{xy}^{i_1 \cdots i_n }=0$, if
\begin{align}
    \mathrm{swap}(M)^\eta |M|^\epsilon m_x m_y
 \prod_i m_i &= -1.
\end{align}
The two atoms are along the $x$-axis. $M$, therefore, does not swap the atoms.
The criterion above then reduces to
\begin{align}
 \prod_i m_i &= -1
\end{align}
Couplings to an odd number of $n^y_\pm$ components are therefore forbidden by the mirror symmetry. Only an even number of $n^y_\pm$ factors is allowed.
Considering instead the $yz$ mirror plane amounts to the redefinition
\begin{equation}
    m_i = \left\lbrace 
    \begin{array}{cc}
        -1,&  \text{if}~ i = x \\
         1,& \text{otherwise}.
    \end{array}\right.
\end{equation}
In this case however, one finds $\mathrm{swap}(M)=-1$. 
And tensor elements for which
\begin{align}
 (-1)^\eta \prod_i m_i &= -1,
\end{align}
are zero. If the tensor is staggered (odd in $\hatn_{-}$) one has $\eta = 1$ and a coupling to an even number of $n_\pm^x$ factors is forbidden.
On the other hand, if the tensor is not staggered (even in $\hatn_{-}$), couplings to an odd number of $n_\pm^x$ is forbidden.
Consider now the case where $n_+^x = 0$. A finite staggered tensor can then only arise from an odd number of $n_{-}^x$ couplings.

\section{Linear perturbation of the Berry curvature}

The goal of this section is to present a derivation of the first order derivative of the Berry curvature as presented in the manuscript in a manifestely gauge-invariant way. 
Consider a continuous partition of unity by a complete set of states:
\begin{equation}
    \id = \sum_{n} \ket{n(\lam)}\bra{n(\lam)}.
\end{equation}
It is parametrized by a vector $\lam$ belonging to the phase space $\mathcal{M}$, representing for example the momentum in the first Brillouin zone and any additional adiabatic real-space coordinates. 
This parametrization is inherited from the Hamiltonian, which can be considered as a map from $\mathcal{M}$ to the space of bounded operators on the Hilbert space $\mathcal{H}$:
\begin{align}
    H\colon &\mathcal{M}\to \mathcal{B}(\mathcal{H})\\
    &\lam \mapsto H_{\lam} .
\end{align}
We consider the more general class of these maps which form the space $C^\infty(\mathcal{M};\mathcal{B}(\mathcal{H}) ) $ of smooth phase space operators.
Given such an operator $\hat{o}$, we can differentiate it with respect to the phase space coordinates.
By simultaneously expanding in terms of the instantaenous basis set, one obtains a notion of covariant differentiation:
\begin{align}
    \partial_i \hat{o}_{\lam} &=
    \partial_i \sum_{nm}  \hat{o}_{\lam}^{nm} \ket{n}\bra{m}
    \notag \\ & =
   \sum_{nm}   ( \partial_i\hat{o}_{\lam}^{nm} ) \ket{n}\bra{m}
   +  \hat{o}_{\lam}^{nm} \big(\partial_i \ket{n}\big)\bra{m}
   +\hat{o}_{\lam}^{nm} \ket{n}\big(\partial_i \bra{m} \big)
   \notag \\ & \equiv \nabla_i  \hat{o}_{\lam} - i [ \Conn_i, \hat{o}_{\lam}] .
\end{align}
Here, we have omitted the explicit notation for the parameter dependence of the states for simplicity.
Further, we introduced the non-abelian Wilzcek-Zee connection
\begin{equation}
    \Conn_i^{nm} \equiv i  \braket{n | \partial_i | m}.
\end{equation}
Using the adjoint representation, this result can be concisely expressed as
\begin{equation}
    \partial_i = \nabla_i - i \mathfrak{A}_i,
\end{equation}
where $\mathfrak{A}_i = \ad~ \Conn_i = [\Conn_i, \bullet] $.
By construction, the associated curvature is flat. 
The reason is that the partial derivatives commute, i.e., one has $[\partial_i, \partial_j] \hat{o}_{\lam} = 0$.
Nevertheless, we can construct it as
\begin{align}
    \mathfrak{F}_{ij} &= i [ \partial_i, \partial_j]
    \notag \\
    & =i [ \nabla_i - i \mathfrak{A}_i, \nabla_j - i \mathfrak{A}_j]
    \notag \\
    & = [ \nabla_i , \mathfrak{A}_j] -  [ \nabla_j , \mathfrak{A}_i] - i  [ \mathfrak{A}_i , \mathfrak{A}_j]
    \notag \\
    & =  \nabla_i  \mathfrak{A}_j -   \nabla_j  \mathfrak{A}_i - i  [ \mathfrak{A}_i , \mathfrak{A}_j] ,
\end{align}
which is the adjoint representation of 
\begin{align}
    \Curv_{ij} = \nabla_i  \Conn_j - \nabla_j \Conn_i - i [\Conn_i, \Conn_j]
= \partial_i  \Conn_j - \partial_j \Conn_i + i [\Conn_i, \Conn_j] = 0 .
\end{align}
Since the right-hand side evaluates to zero for the flat connection, one has the two identities
\begin{align}
    \nabla_i  \Conn_j - \nabla_j \Conn_i & =  +i [\Conn_i, \Conn_j]
    \\
    \partial_i  \Conn_j - \partial_j \Conn_i & = -  i [\Conn_i, \Conn_j].
    \label{eq:non_abelian_curvature}
\end{align}
Let $\ket{n(\lam)}$ now represent the parameter-dependent eigenstates of the Hamiltonian. 
The abelian Berry curvature tensor of the $n$-th band is then given by
 \begin{equation}
     \Omega_{\mu \nu}^{n}(\lam)=i \sum_{m \neq n}
    \frac{(\partial_\mu H)^{nm} (\partial_\nu H)^{mn}}{(\epsilon_n - \epsilon_m)^2} - ( \mu\leftrightarrow\nu ).
 \end{equation}
 Since one has $(\partial_\mu H)^{nm} = i (\epsilon_n-\epsilon_m) \Conn_\mu^{nm}$ for $n\neq m$, this can also be written as
 \begin{align}
     \Omega_{\mu \nu}^{n}(\lam)=  - i [\Conn_\mu, \Conn_\nu]^{nn} =  (\partial_\mu  \Conn_\nu - \partial_\nu \Conn_\mu)^{nn}\equiv \Omega_{\mu \nu}^{nn}(\lam) ,
 \end{align}
 which is therefore in accordance with Eq.~(\ref{eq:non_abelian_curvature}).
 We can take this as the definition of the abelian part of the more general, but flat curvature tensor:
 \begin{equation}
     \Curv_{\mu\nu} = \Omega_{\mu \nu} + i [\Conn_\mu, \Conn_\nu] = 0.
 \end{equation}
 For the calculation of the anomalous Hall effect, one weights the Berry curvature of the respective bands with the electronic density matrix $\rho$, from which one obtains the total Berry curvature
 \begin{align}
    \Omega_{\mu \nu}^\mathrm{tot}(\lam) = -i ~\tr ~\rho  [\Conn_\mu, \Conn_\nu] = \Im~ \tr ~\rho  [\Conn_\mu, \Conn_\nu],
 \end{align}
 and where the density matrix is given by
 \begin{equation}
     \rho(\lam) = \sum_{n}  n_\mathrm{F}(\epsilon_n) \ket{n}\bra{n},
 \end{equation}
 where $ n_\mathrm{F}$ is the Fermi-Dirac distribution at the chemical potential $\mu$.
 Having introduced the necessary terminology, the first order derivation of the Berry curvature tensor can now be constructed.
 Several terms can contribute:
 \begin{align}
    \partial_\kappa \Omega_{\mu \nu}^\mathrm{tot} & = 
    ( \Im~ \tr ~\partial_\kappa \rho  \Conn_\mu \Conn_\nu
    + \Im~ \tr ~\rho  \partial_\kappa \Conn_\mu \Conn_\nu
   + \Im~ \tr ~\rho  \Conn_\mu \partial_\kappa \Conn_\nu
    ) - ( \mu \leftrightarrow \nu)
\notag \\
& = \Im ~\tr \left(
\nabla_\kappa \rho \Conn_\mu \Conn_\nu - i [\Conn_\kappa, \rho ] \Conn_\mu \Conn_\nu + \rho [ \partial_\kappa \Conn_\mu, \Conn_\nu]
\right) - ( \mu \leftrightarrow \nu) .
 \end{align}
As $\Omega_{\mu \nu}^\mathrm{tot}$ is gauge-invariant quantity, its derivative is gauge-invariant as well, and so is the right-hand side of this equation.
 The first term is a Fermi surface contribution which we single out:
 \begin{equation}
    (\partial_\kappa \Omega_{\mu \nu}^\mathrm{tot})^\mathrm{FS} = \Im ~\tr 
\nabla_\kappa \rho [\Conn_\mu, \Conn_\nu] .
 \end{equation}
 It vanishes for insulators in the zero temperature limit and we neglect it in the following.
 The remaining terms are 
 \begin{align}
    (\partial_\kappa \Omega_{\mu \nu}^\mathrm{tot})
& = \Im ~\tr \rho \left(
-i [\Conn_\mu \Conn_\nu, \Conn_\kappa ] + [\partial_\kappa \Conn_\mu , \Conn_\nu]
\right) - ( \mu \leftrightarrow \nu)
\notag \\
& = \frac{1}{2}\Im ~\tr \rho \left(
 [\Omega_{\mu\nu}, \Conn_\kappa ]
 +[\mathcal{Q}_{\kappa\mu}, \Conn_\nu ]
 +[\Omega_{\kappa\mu}, \Conn_\nu ]
\right) - ( \mu \leftrightarrow \nu)
\notag \\
& = \frac{1}{2}\Im ~\tr \rho \left(
 [\Omega_{\mu\nu}, \Conn_\kappa ]
 +[\mathcal{Q}_{\kappa\mu}, \Conn_\nu ]
 +[\Omega_{\nu\kappa}, \Conn_\mu ]
\right) - ( \mu \leftrightarrow \nu),
 \end{align}
 where we have introduced
 \begin{equation}
     \mathcal{Q}_{\mu\nu} = 
     \partial_\mu  \Conn_\nu + \partial_\nu \Conn_\mu ,
 \end{equation}
 which is related to the quantum metric tensor.